\documentclass[reprint,twocolumn,aps,prl,floats,superscriptaddress]{revtex4-1}
\usepackage[english]{babel}
\usepackage[ansinew]{inputenc}
\usepackage{amsmath}

\usepackage{amssymb}
\usepackage{dsfont}
\usepackage{bbold}
\usepackage{lineno}
\usepackage{braket}
\usepackage{color,soul}
\usepackage{chngcntr}
\usepackage{color}
\usepackage{dcolumn}
\usepackage{textcomp}

\usepackage{graphicx}
\usepackage{hyperref}
\hypersetup{
	colorlinks=true,
	linkcolor=blue,
	filecolor=magenta,
	citecolor=red,
	urlcolor=cyan,
	bookmarks=true
}

\begin{document}
\title{Spin excitations in a $4f-3d$ heterodimer on MgO}
\author{A.~Singha}
\affiliation{Center for Quantum Nanoscience, Institute for Basic Science (IBS), Seoul 03760, Republic of Korea}
\affiliation{Institute of Physics, {\'E}cole Polytechnique F{\'e}d{\'e}rale de Lausanne, Station 3, CH-1015 Lausanne, Switzerland}
\affiliation{Department of Physics, Ewha Womans University, Seoul 03760, Republic of Korea}
\author{F.~Donati}
\affiliation{Center for Quantum Nanoscience, Institute for Basic Science (IBS), Seoul 03760, Republic of Korea}
\affiliation{Institute of Physics, {\'E}cole Polytechnique F{\'e}d{\'e}rale de Lausanne, Station 3, CH-1015 Lausanne, Switzerland}
\affiliation{Department of Physics, Ewha Womans University, Seoul 03760, Republic of Korea}
\author{F.~D.~Natterer}
\affiliation{Institute of Physics, {\'E}cole Polytechnique F{\'e}d{\'e}rale de Lausanne, Station 3, CH-1015 Lausanne, Switzerland}
\author{C.~W\"ackerlin}
\affiliation{Institute of Physics, {\'E}cole Polytechnique F{\'e}d{\'e}rale de Lausanne, Station 3, CH-1015 Lausanne, Switzerland}
\affiliation{Nanoscale Materials Science, Empa, Swiss Federal Laboratories for Materials Science and Technology, 8600 D{\"u}bendorf, Switzerland}
\author{S.~Stavri\'c}
\affiliation{Vin\v{c}a Institute of Nuclear Sciences, University of Belgrade, Serbia}
\author{Z.~S.~Popovi\'c}
\affiliation{Vin\v{c}a Institute of Nuclear Sciences, University of Belgrade, Serbia}
\author{{\v Z}. {\v S}ljivan{\v c}anin}
\affiliation{Vin\v{c}a Institute of Nuclear Sciences, University of Belgrade, Serbia}
\affiliation{Texas A\&M University at Qatar, Doha, Qatar}
\author{F.~Patthey}
\affiliation{Institute of Physics, {\'E}cole Polytechnique F{\'e}d{\'e}rale de Lausanne, Station 3, CH-1015 Lausanne, Switzerland}
\author{H.~Brune}
\affiliation{Institute of Physics, {\'E}cole Polytechnique F{\'e}d{\'e}rale de Lausanne, Station 3, CH-1015 Lausanne, Switzerland}

\begin{abstract}
We report on the magnetic properties of HoCo dimers as a model system for the smallest intermetallic transition metal-lanthanide compound. The dimers are adsorbed on ultrathin MgO(100) films grown on Ag(100). New for $4f$ elements, we detect inelastic excitations with scanning tunneling microscopy and prove by their behaviour in applied magnetic field that they are spin-excitations. In combination with density functional theory and spin Hamiltonian analysis we determine the magnetic level distribution, as well as sign and magnitude of the exchange interaction between the two atoms. In contrast to typical $4f-3d$ bulk compounds, we find ferromagnetic coupling in the dimer.
\end{abstract}
\maketitle

Many alloys combining transition metal (TM) elements of the first row with rare earth (RE) elements, are widely used as permanent magnets due to their large magnetic anisotropy. On a more fundamental level, the coupling between the spins of these elements can give rise to complex magnetic structures that can exhibit rich phase diagrams~\cite{gub73}. This is due to the indirect exchange interaction between the $4f$ orbitals of the RE and the $3d$ orbitals of the TM mediated by the $spd$ conduction electrons~\cite{cam72}. In addition, the magnetic order of RE-TM alloys is strongly affected by structural relaxations and surface effects~\cite{bac93}. Both become particularly important when the size of the magnet reaches atomic dimensions, as demonstrated for several single molecule magnets (SMM)~\cite{mis04, dre12, wes12, vie13, wes15}.

Understanding the magnetic level splitting and therefore the origin of magnetic anisotropy is required for the rational design of prototypical nanomagnets. The splitting of the low energy levels can be investigated using scanning tunneling microscopy (STM) and spin-excitation spectroscopy (SES)~\cite{hei04}. However, the detection of spin-excitations in $4f$ atoms and $4f$-SMMs is very challenging due to the vanishing or very small contribution of the $f$-density of states at the Fermi level~\cite{cof15, ste15}, and reading the magnetic state of RE atoms and islands is possibly enabled due to the exchange interaction with the $5d$ shell~\cite{cof15, nat17, mat98}. On the other hand, the $3d$ orbitals of TM atoms are more directly probed by tunneling electrons and can show large SES cross-sections~\cite{rau14}. Therefore we investigate the smallest surface-supported RE-TM alloy, namely a heterodimer, as a model system for investigating the $4f-3d$ exchange coupling at the atomic scale.

We report on the magnetic properties of HoCo heterodimers on MgO(100) thin films grown on Ag(100). The properties of Ho and Co on MgO are particularly intriguing: Ho on MgO is the first discovered single atom magnet exhibiting the highest possible magnetic moment~\cite{don16, nat17}, whereas Co on the same surface exhibits the largest magnetic anisotropy energy~\cite{rau14}. The MgO(100) thin films grown on metal surfaces are ideal substrates as MgO decouples the magnetic states of adsorbates from the scattering of substrate electrons and soft phonons~\cite{don16, wac16}. For HoCo dimers, we observe two pairs of spin-excitations at $\pm 8$~meV and $\pm 20$~meV, for which we develop an effective spin Hamiltonian (SH), using minimal input from our density functional theory (DFT) calculations. The SH model reproduces the magnetic field dependent conductance steps. In agreement with DFT, this model finds collinear ferromagnetic coupling between Ho and Co. The DFT inferred adsorption site is also in agreement with experiment. We determine the HoCo magnetic level spectrum and the relative contribution of the two elements to the experimentally detected spin-excitations.

\begin{figure}
\begin{center}
\includegraphics{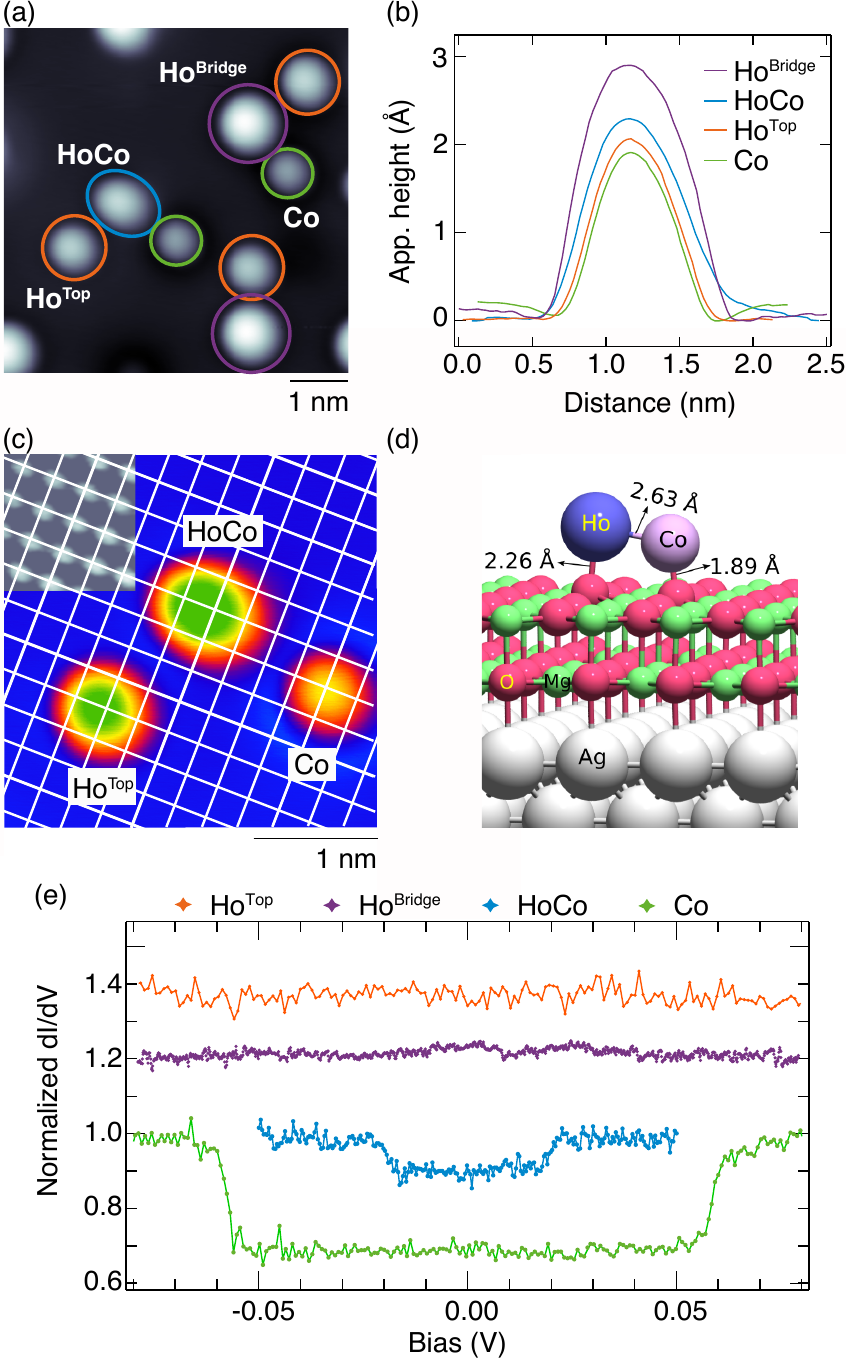}
\end{center}
\caption{Identification of HoCo dimers and detection of their spin-excitations. (a) STM image and (b) apparent height profiles of HoCo dimer and individual Ho$^{\rm top}$, Ho$^{\rm br}$, and Co atoms. (c) STM image showing the characteristic elliptical shape of a HoCo dimer in comparison with the more circular individual Ho and Co atoms ((a) - (c) $V_{\rm t} = 100$~mV, $I_{\rm t} = 50$~pA, and $\mu_{0} H = 6$~T, $T = 4.3$~K). Inset: atomic resolution image of 2~ML MgO with O imaged bright ($V_{\rm t} = 10$~mV, $I_{\rm t} = 8$~nA). The white grid shows the O sublattice~\cite{fer17}. (d) DFT calculated adsorption geometry of HoCo. (e) $dI/dV$ of HoCo dimer ($V_{\rm t} = 40$~mV, $I_{\rm t} = 250$~pA), Ho$^{\rm top}$, Ho$^{\rm br}$, and Co atoms ($V_{\rm t} = 100$~mV, $I_{\rm t} = 500$~pA). All spectra: $V_{\rm mod, ptp} = 1$~mV, $\mu_{0}H = 1$~T, $T = 4.3$~K. For clarity, Ho$^{\rm top}$ and Ho$^{\rm br}$ spectra are vertically offset by 0.2 each.}
\label{id}
\end{figure}

The Ho and Co were dosed onto the cold substrates simultaneously, in the measurement position of our $0.4$~K, $\pm8$~T STM home-built STM~\cite{cla05}. This yields predominantly individual Ho and Co atoms, but also the occasional formation of homo and heterodimers. In order to unequivocally distinguish the different species, we separately prepared samples with only Ho and only Co. Each species has characteristic inelastic conductance ($dI/dV$) steps and/or apparent heights. The homodimers Ho$_2$ and Co$_2$ show intense $dI/dV$ steps located at $\pm 85$ and $\pm 13$~meV, respectively (Figures \ref{figS1} and \ref{figS2}). Figure~\ref{id}(a) shows an STM image of the remaining four species. Isolated Ho atoms adsorb on-top of O (Ho$^{\rm top}$) or bridge sites (Ho$^{\rm br}$) of the MgO(100) lattice~\cite{fer17}. They are discerned by their distinct apparent heights (Fig.~\ref{id}(b)); neither one has observable inelastic conductance steps (Fig.~\ref{id}(e)). Isolated Co atoms adsorb on-top of O only~\cite{fer17} and are clearly identified by their $dI/dV$ steps at $\pm 58$~meV, reminiscent of their high magnetic anisotropy~\cite{rau14}. The apparent height of the HoCo-dimer is distinct from Ho$^{\rm br}$, Ho$^{\rm top}$, and Co atoms (Fig.~\ref{id}(b)), and it possesses an "egg-like" footprint with its axis aligned along the MgO(100) surface lattice directions (Fig.~\ref{id}(c)). This shape, as well as the location of the dimer, are in agreement with the adsorption geometry inferred from DFT, where Ho and Co adsorb on two adjacent O sites with vertical distances of 2.26 and 1.89~{\AA} (Fig.~\ref{id}(d)). 

\begin{figure}[b!]
\begin{center}
\includegraphics{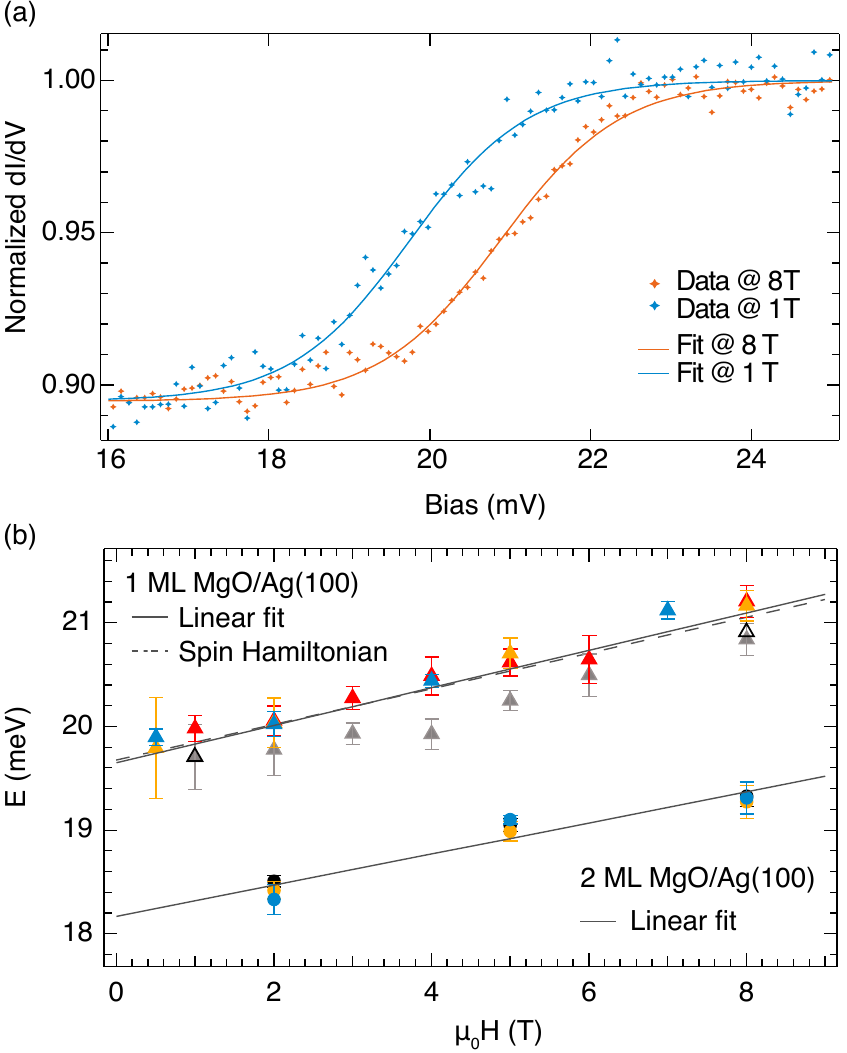}
\end{center}
\caption{(a) Magnetic field dependence of the $dI/dV$ step of a HoCo dimer (dots: raw data; lines: sigmoid fits). Each data point represents the average from 10 spectra recorded on the same HoCo dimer ($V_{\rm t} = 30$~mV, $I_{\rm t} = 750$~pA, $V_{\rm mod, ptp} = 500$~$\mu$V, $T = 0.4$~K). (b) Zeeman plot obtained from measurements on four HoCo dimers adsorbed on 1~ML MgO/Ag(100) (filled triangles) and three HoCo dimers on 2~ML MgO/Ag(100) (filled circles) ($T = 4$~K). Different colors represent the individual heterodimers. The 0.4~K measurements shown in (a) are included as open black triangles. Error bars represent the standard deviation from $\geq 5$ measurements on the same heterodimer. The dashed line shows a fit with the spin Hamiltonian described in the text.}
\label{SES}
\end{figure}
The spectroscopic fingerprints of the heterodimer are conductance steps at $\pm 20$~meV (Fig.~\ref{id}(e)). Their magnetic origin becomes evident from their linear shift in an out-of-plane external magnetic field (Fig.~\ref{SES}(a)). Fig.~\ref{SES}(b) displays the magnetic field and MgO thickness-dependent excitation energies for several HoCo dimers. From a linear fit to the step energy $E(H) = g \, \mu_{0} H \, \Delta m \, \mu_{\rm B}$  (Fig.~\ref{SES}(b)), we extract the effective Land\'e $g-$factor of $3.1 \pm 0.3$ for the observed transition (Table~\ref{tabS1}), assuming the change in the out-of-plane component of total magnetic moment to be $\Delta m = \pm 1$. The large mean value of $g$ indicates the presence of a large orbital moment, as was previously reported for Fe~\cite{bau15}.  

As can be seen from Fig.~\ref{SES}(b), the step energy depends strongly on MgO thickness ({\it i.e.}, it moves to lower energy by $1.7$~meV for thicker MgO layer), and also weakly on the local environment of a given dimer (variance of $0.2$~meV for the dimer shown with gray triangles). According to DFT, the adsorption of the HoCo dimers leads to sizeable distortions of the underlying MgO lattice. These distortions are different for 1 and 2~ML MgO/Ag(100), thus creating different crystal fields. Similar to the case of Fe atoms on MgO, HoCo dimers show lower excitation energies when adsorbed on thicker MgO films~\cite{pau16}.

\begin{figure}[b!]
\begin{center}
\includegraphics{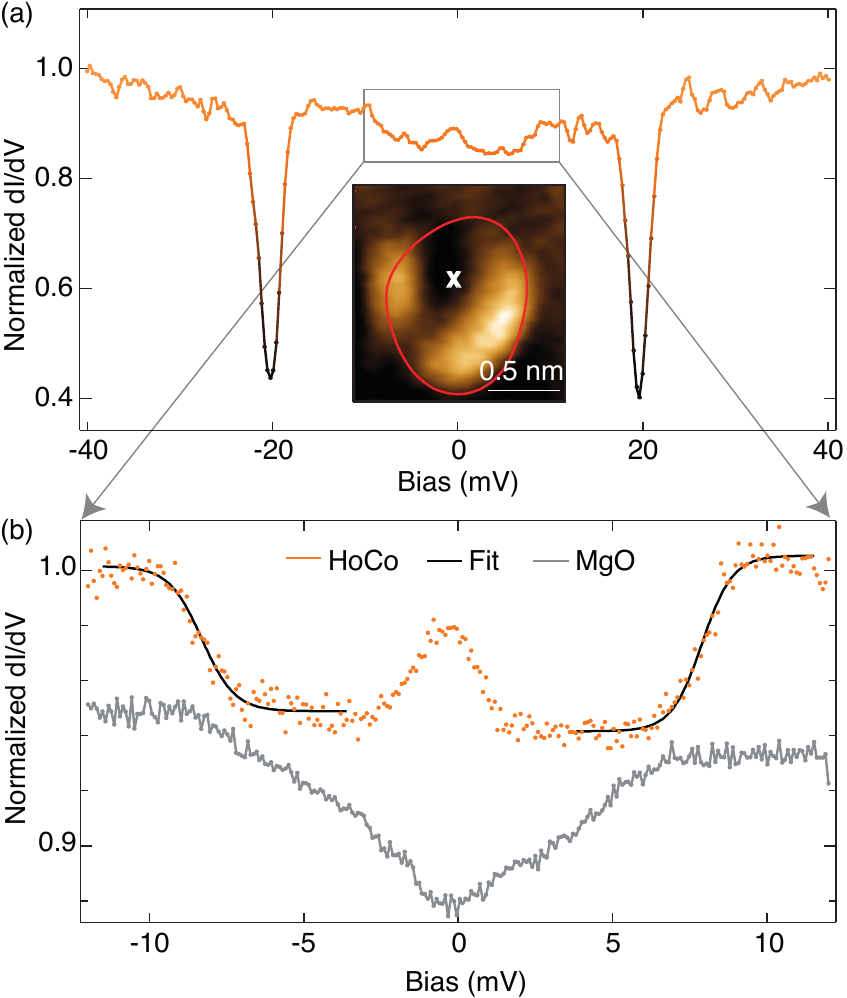}
\end{center}
\caption{(a) $dI/dV$ spectra measured with a spin-polarized tip at the center of a HoCo dimer (dots: measurements, lines: sigmoid fits). Inset shows the corresponding $dI/dV$ map measured at $V_{\rm t} = 20$~mV overlaid with the contour of 15$\%$ of the maximum apparent height of the heterodimer. In addition to the jagged inelastic features at $\pm 20$~meV, a pair of inner steps are detected at about $\pm 8$~meV ($V_{\rm t} = 40$~mV, $I_{\rm t} = 1$~nA, and $V_{\rm mod, ptp} = 500$~$\mu$V). (b) Zoom on the inner step shown in (a). Each data point represents the average from 5 acquisitions. ($V_{\rm t} = 15$~mV, $I_{\rm t} = 300$~pA, and $V_{\rm mod, ptp} = 200$~$\mu$V). $\mu_{0}H = 8$~T and $T = 0.4$~K for both figures.}
\label{SP-tip}
\end{figure}

We expect two types of magnetic excitations in dimers, one where the total magnetic moment $\bf \hat S$ changes and one where its projection $\rm \hat S_{\rm z}$ does~\cite{sch10}. The detection of only one transition suggests that the intensity of the second is very low, and possibly obscured by noise. We therefore used spin-polarized (SP) STM tips that can enhance the cross-section of a spin-excitation processes allowing to probe very weak SES transitions~\cite{bau15b, bau15, lot10}. We spin polarized the current by transferring Co atoms to the tip until a signature of SP in the SES of Co was observed~\cite{bau15}. Figure~\ref{SP-tip}(a) shows that the steps at $\pm 20$~meV are now intense dips and indeed, an additional pair of symmetric steps at $\pm 8.1$~meV becomes apparent (Fig.~\ref{SP-tip}(b)). Being only 13~\% of the $\pm 20$~meV step conductance, the transition would be only $1\%$ without SP and thus obscured by noise in Fig.~\ref{id}(e). Note that the low magnetic field SP measurement of the $\pm8$~meV steps were not possible in the present setup.

The observed inelastic features are mapped onto an effective spin Hamiltonian (SH) of the following form: $\hat H = D \hat S_{z}^{2} + J ({\bf \hat S}_{\rm Ho} \cdot {\bf \hat S}_{\rm Co} ) + \mu_{\rm B} [g_{\rm Ho} {\bf \hat S}_{\rm Ho}+ g_{\rm Co} {\bf \hat S}_{\rm Co}] \cdot \mu_{0} \bf H$, where $D$ is the uniaxial out-of-plane ($z$) anisotropy, $J$ the Heisenberg exchange coupling between effective Ho and Co spin, and the last term is the Zeeman energy due to the external field acting on both effective spins. $\hat H$ includes the effective $g$ factors of the individual atoms (see below). 

The effective spin values define the level multiplicity of the lowest multiplet. For REs the spin-orbit coupling largely dominates over the crystal field and, therefore, the effective spin can be well described using the total magnetic moment (orbital + spin)~\cite{jan15}. Accordingly, for Ho we consider the highest possible projection for the total magnetic moment in $4f^{11}$ configuration as found in DFT, yielding $g_{\rm Ho} = 1.2$ and $S_{\rm Ho} = 15/2$. In contrast, for TMs the hierarchy of interactions is reversed. As a result of this, the total moment is not any more a good quantum number~\cite{bau15, rau14}. In particular for low symmetry environments, the level multiplicity can be defined by the spin moment only~\cite{hir07, ott08}. Therefore, for Co, we take the spin magnetic moment $S_{\rm Co} = 1$ calculated from DFT, and include the possibility of a non-vanishing orbital component through $g_{\rm Co}$. This leaves $D$, $J$, and $g_{\rm Co}$ as the only free parameters. Note that the Ho-$4f$ and Co-$3d$ occupancy in the dimer differ from the respective single atom counterparts~\cite{rau14, don16}.

Both inelastic steps, as well as the large effective $g$ factor for the spin-excitation at $\pm 20$~meV, are reproduced with $D = -0.47$~meV (out-of-plane easy axis), ferromagnetic coupling with $J = -1.43$~meV, and $g_{\rm Co} = 3.2$. Figure~\ref{SES}(b) illustrates the excellent agreement between measured (full line) and calculated (dashed line) Zeeman shift of the step energy. A positive $J$ would yield $g < 2$, which contradicts our experimental observation of $g=3.1\pm0.3$. The negative value of $J$ indicates ferromagnetic coupling between the two atoms in the heterodimer as also found in DFT. Note that most bulk RE-TM alloys have ferrimagnetic exchange~\cite{jan03}.

\begin{figure}[h!]
\begin{center}
\includegraphics{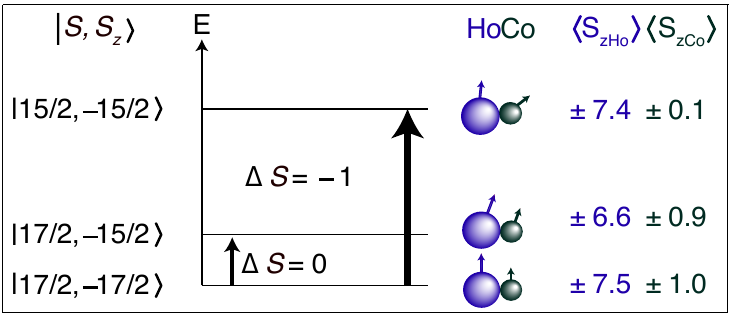}
\end{center}
\caption{Schematic of the magnetic sub-levels involved in the SES process. The $\Delta S = -1$ transition (thick arrow) causes the high energy $dI/dV$ steps and implies a large change in $S_{z, \rm Co}$ and therefore has large signal. The low energy steps are due to a $\Delta S = 0$ transition, where $S_{z, \rm Co}$ changes only slightly (thin arrow).}
\label{levels}
\end{figure}

The selection rules for inelastic spin-excitations mediated by tunnel electrons are $\Delta S = \pm 1, 0$, the same applies to the $z$-projection of total dimer moment $S$. Figure~\ref{levels} shows the ground state and the first two excited states which are accessible following these selection rules. The two transitions observed in experiment are $\ket{S, S_z} = \ket{17/2, \pm 17/2} \rightarrow \ket{15/2, \pm15/2}$, for which we calculate 21.1~meV at 8~T, and $\ket{17/2, \pm 17/2} \rightarrow \ket{17/2, \pm 15/2}$, for which our SH yields 8.2~meV at 8~T. 

In order to quantify the contribution of each of the atoms to $S_{z}$, we compute the expectation value of $S_{z, \rm Ho}$ and $S_{z, \rm Co}$ (supplementary information) and show them on the right-hand side of Fig.~\ref{levels}. The prominent inelastic transition observed at $\pm 20$~meV is dominated by an almost 90~\% reduction of $S_{z, \rm Co}$, whereas the out-of-plane projection of the Ho moment is reduced only by a few percent. If we assumed that the tunnel electrons interacted mostly with the TM element, this would explain the larger step height for this transition implying a strong change of $S_{z, \rm Co}$. For the inner steps at $\pm 8$~meV the opposite is true, \textit{i.e.}, the change of the out-of-plane projected overall magnetic moment of the dimer is mostly due to the Ho atom, again in agreement with the observation of these steps having low intensity. We finally attribute the large $g_{\rm Co} = 3.2$ value to an unquenched orbital moment contribution of $m_L = 1.2~\mu_{\rm B}$. This value compares well with previously reported values for single surface-supported Co atoms~\cite{gam03, don14}. However, it is strongly reduced compared to the maximum orbital moment of Co atoms on MgO~\cite{rau14}, due to the lowered symmetry caused by the neighbouring Ho atom in the heterodimer.

In conclusion, we demonstrated the detection of spin-excitations of individual exchange-coupled RE-TM heterodimers. The spectroscopic cross-section mostly results from the out-of-plane projected spin variations in the $3d$ element. The possibility of detecting inelastic transitions of magnetic origin in RE based nanostructures enables to unravel their magnetic level spectrum, as well as their internal magnetic coupling.

\begin{acknowledgements}
We acknowledge funding from the Swiss National Science Foundation (SNSF) through grants 200020--157081 (AS), PZ00P2--167965 (FDN), PZ00P2--142474 (CW), and IZ73Z0--152406 (SS). SS, ZSP, and {\v Z}{\v S} acknowledge funding from the Serbian Ministry of Education and Science (OI--171033). The DFT calculations were performed at UK (Archer) and Swedish (Beskow) supercomputing facilities, available through PRACE DECI--13.
\end{acknowledgements}

%


\begin{thebibliography}{41}%
	\makeatletter
	\providecommand \@ifxundefined [1]{%
		\@ifx{#1\undefined}
	}%
	\providecommand \@ifnum [1]{%
		\ifnum #1\expandafter \@firstoftwo
		\else \expandafter \@secondoftwo
		\fi
	}%
	\providecommand \@ifx [1]{%
		\ifx #1\expandafter \@firstoftwo
		\else \expandafter \@secondoftwo
		\fi
	}%
	\providecommand \natexlab [1]{#1}%
	\providecommand \enquote  [1]{``#1''}%
	\providecommand \bibnamefont  [1]{#1}%
	\providecommand \bibfnamefont [1]{#1}%
	\providecommand \citenamefont [1]{#1}%
	\providecommand \href@noop [0]{\@secondoftwo}%
	\providecommand \href [0]{\begingroup \@sanitize@url \@href}%
	\providecommand \@href[1]{\@@startlink{#1}\@@href}%
	\providecommand \@@href[1]{\endgroup#1\@@endlink}%
	\providecommand \@sanitize@url [0]{\catcode `\\12\catcode `\$12\catcode
		`\&12\catcode `\#12\catcode `\^12\catcode `\_12\catcode `\%12\relax}%
	\providecommand \@@startlink[1]{}%
	\providecommand \@@endlink[0]{}%
	\providecommand \url  [0]{\begingroup\@sanitize@url \@url }%
	\providecommand \@url [1]{\endgroup\@href {#1}{\urlprefix }}%
	\providecommand \urlprefix  [0]{URL }%
	\providecommand \Eprint [0]{\href }%
	\providecommand \doibase [0]{http://dx.doi.org/}%
	\providecommand \selectlanguage [0]{\@gobble}%
	\providecommand \bibinfo  [0]{\@secondoftwo}%
	\providecommand \bibfield  [0]{\@secondoftwo}%
	\providecommand \translation [1]{[#1]}%
	\providecommand \BibitemOpen [0]{}%
	\providecommand \bibitemStop [0]{}%
	\providecommand \bibitemNoStop [0]{.\EOS\space}%
	\providecommand \EOS [0]{\spacefactor3000\relax}%
	\providecommand \BibitemShut  [1]{\csname bibitem#1\endcsname}%
	\let\auto@bib@innerbib\@empty
	\bibitem [{\citenamefont {Gubbens}\ and\ \citenamefont
		{Buschow}(1973)}]{gub73}%
	\BibitemOpen
	\bibfield  {author} {\bibinfo {author} {\bibfnamefont {P.~C.~M.}\
			\bibnamefont {Gubbens}}\ and\ \bibinfo {author} {\bibfnamefont {K.~H.~J.}\
			\bibnamefont {Buschow}},\ }\href {\doibase 10.1063/1.1662832} {\bibfield
		{journal} {\bibinfo  {journal} {J. Appl. Phys.}\ }\textbf {\bibinfo {volume}
			{44}},\ \bibinfo {pages} {3739} (\bibinfo {year} {1973})}\BibitemShut
	{NoStop}%
	\bibitem [{\citenamefont {Campbell}(1972)}]{cam72}%
	\BibitemOpen
	\bibfield  {author} {\bibinfo {author} {\bibfnamefont {I.~A.}\ \bibnamefont
			{Campbell}},\ }\href {http://stacks.iop.org/0305-4608/2/i=3/a=004} {\bibfield
		{journal} {\bibinfo  {journal} {J. Phys. F: Metal Physics}\ }\textbf
		{\bibinfo {volume} {2}},\ \bibinfo {pages} {L47} (\bibinfo {year}
		{1972})}\BibitemShut {NoStop}%
	\bibitem [{\citenamefont {Baczewski}\ \emph {et~al.}(1993)\citenamefont
		{Baczewski}, \citenamefont {Givord}, \citenamefont {Alameda}, \citenamefont
		{Dieny}, \citenamefont {Nozieres}, \citenamefont {Rebouillat},\ and\
		\citenamefont {Prejean}}]{bac93}%
	\BibitemOpen
	\bibfield  {author} {\bibinfo {author} {\bibfnamefont {L.~T.}\ \bibnamefont
			{Baczewski}}, \bibinfo {author} {\bibfnamefont {D.}~\bibnamefont {Givord}},
		\bibinfo {author} {\bibfnamefont {J.~M.}\ \bibnamefont {Alameda}}, \bibinfo
		{author} {\bibfnamefont {B.}~\bibnamefont {Dieny}}, \bibinfo {author}
		{\bibfnamefont {J.}~\bibnamefont {Nozieres}}, \bibinfo {author}
		{\bibfnamefont {J.~P.}\ \bibnamefont {Rebouillat}}, \ and\ \bibinfo {author}
		{\bibfnamefont {J.~J.}\ \bibnamefont {Prejean}},\ }\href {\doibase
		10.12693/APhysPolA.83.629} {\bibfield  {journal} {\bibinfo  {journal} {Acta
				Phys. Pol. A.}\ }\textbf {\bibinfo {volume} {83}},\ \bibinfo {pages} {629}
		(\bibinfo {year} {1993})}\BibitemShut {NoStop}%
	\bibitem [{\citenamefont {Mishra}\ \emph {et~al.}(2004)\citenamefont {Mishra},
		\citenamefont {Wernsdorfer}, \citenamefont {Abboud},\ and\ \citenamefont
		{Christou}}]{mis04}%
	\BibitemOpen
	\bibfield  {author} {\bibinfo {author} {\bibfnamefont {A.}~\bibnamefont
			{Mishra}}, \bibinfo {author} {\bibfnamefont {W.}~\bibnamefont {Wernsdorfer}},
		\bibinfo {author} {\bibfnamefont {K.~A.}\ \bibnamefont {Abboud}}, \ and\
		\bibinfo {author} {\bibfnamefont {G.}~\bibnamefont {Christou}},\ }\href
	{\doibase 10.1021/ja0452727} {\bibfield  {journal} {\bibinfo  {journal} {J.
				Am. Chem. Soc.}\ }\textbf {\bibinfo {volume} {126}},\ \bibinfo {pages}
		{15648} (\bibinfo {year} {2004})}\BibitemShut {NoStop}%
	\bibitem [{\citenamefont {Dreiser}\ \emph {et~al.}(2012)\citenamefont
		{Dreiser}, \citenamefont {Pedersen}, \citenamefont {Birk}, \citenamefont
		{Schau-Magnussen}, \citenamefont {Piamonteze}, \citenamefont {Rusponi},
		\citenamefont {Weyherm\"uller}, \citenamefont {Brune}, \citenamefont
		{Nolting},\ and\ \citenamefont {Bendix}}]{dre12}%
	\BibitemOpen
	\bibfield  {author} {\bibinfo {author} {\bibfnamefont {J.}~\bibnamefont
			{Dreiser}}, \bibinfo {author} {\bibfnamefont {K.~S.}\ \bibnamefont
			{Pedersen}}, \bibinfo {author} {\bibfnamefont {T.}~\bibnamefont {Birk}},
		\bibinfo {author} {\bibfnamefont {M.}~\bibnamefont {Schau-Magnussen}},
		\bibinfo {author} {\bibfnamefont {C.}~\bibnamefont {Piamonteze}}, \bibinfo
		{author} {\bibfnamefont {S.}~\bibnamefont {Rusponi}}, \bibinfo {author}
		{\bibfnamefont {T.}~\bibnamefont {Weyherm\"uller}}, \bibinfo {author}
		{\bibfnamefont {H.}~\bibnamefont {Brune}}, \bibinfo {author} {\bibfnamefont
			{F.}~\bibnamefont {Nolting}}, \ and\ \bibinfo {author} {\bibfnamefont
			{J.}~\bibnamefont {Bendix}},\ }\href {\doibase 10.1021/jp303512a} {\bibfield
		{journal} {\bibinfo  {journal} {J. Phys. Chem. A}\ }\textbf {\bibinfo
			{volume} {116}},\ \bibinfo {pages} {7842} (\bibinfo {year}
		{2012})}\BibitemShut {NoStop}%
	\bibitem [{\citenamefont {Westerstr\"om}\ \emph {et~al.}(2012)\citenamefont
		{Westerstr\"om}, \citenamefont {Dreiser}, \citenamefont {Piamonteze},
		\citenamefont {Muntwiler}, \citenamefont {Weyeneth}, \citenamefont {Brune},
		\citenamefont {Rusponi}, \citenamefont {Nolting}, \citenamefont {Popov},
		\citenamefont {Yang}, \citenamefont {Dunsch},\ and\ \citenamefont
		{Greber}}]{wes12}%
	\BibitemOpen
	\bibfield  {author} {\bibinfo {author} {\bibfnamefont {R.}~\bibnamefont
			{Westerstr\"om}}, \bibinfo {author} {\bibfnamefont {J.}~\bibnamefont
			{Dreiser}}, \bibinfo {author} {\bibfnamefont {C.}~\bibnamefont {Piamonteze}},
		\bibinfo {author} {\bibfnamefont {M.}~\bibnamefont {Muntwiler}}, \bibinfo
		{author} {\bibfnamefont {S.}~\bibnamefont {Weyeneth}}, \bibinfo {author}
		{\bibfnamefont {H.}~\bibnamefont {Brune}}, \bibinfo {author} {\bibfnamefont
			{S.}~\bibnamefont {Rusponi}}, \bibinfo {author} {\bibfnamefont
			{F.}~\bibnamefont {Nolting}}, \bibinfo {author} {\bibfnamefont
			{A.}~\bibnamefont {Popov}}, \bibinfo {author} {\bibfnamefont
			{S.}~\bibnamefont {Yang}}, \bibinfo {author} {\bibfnamefont {L.}~\bibnamefont
			{Dunsch}}, \ and\ \bibinfo {author} {\bibfnamefont {T.}~\bibnamefont
			{Greber}},\ }\href {\doibase 10.1021/ja301044p} {\bibfield  {journal}
		{\bibinfo  {journal} {J. Am. Chem. Soc.}\ }\textbf {\bibinfo {volume}
			{134}},\ \bibinfo {pages} {9840} (\bibinfo {year} {2012})}\BibitemShut
	{NoStop}%
	\bibitem [{\citenamefont {Vieru}\ \emph {et~al.}(2013)\citenamefont {Vieru},
		\citenamefont {Ungur},\ and\ \citenamefont {Chibotaru}}]{vie13}%
	\BibitemOpen
	\bibfield  {author} {\bibinfo {author} {\bibfnamefont {V.}~\bibnamefont
			{Vieru}}, \bibinfo {author} {\bibfnamefont {L.}~\bibnamefont {Ungur}}, \ and\
		\bibinfo {author} {\bibfnamefont {L.~F.}\ \bibnamefont {Chibotaru}},\ }\href
	{\doibase 10.1021/jz4017206} {\bibfield  {journal} {\bibinfo  {journal} {J.
				Phys. Chem. Lett.}\ }\textbf {\bibinfo {volume} {4}},\ \bibinfo {pages}
		{3565} (\bibinfo {year} {2013})}\BibitemShut {NoStop}%
	\bibitem [{\citenamefont {Westerstr\"om}\ \emph {et~al.}(2015)\citenamefont
		{Westerstr\"om}, \citenamefont {Uldry}, \citenamefont {Stania}, \citenamefont
		{Dreiser}, \citenamefont {Piamonteze}, \citenamefont {Muntwiler},
		\citenamefont {Matsui}, \citenamefont {Rusponi}, \citenamefont {Brune},
		\citenamefont {Yang}, \citenamefont {Popov}, \citenamefont {B\"uchner},
		\citenamefont {Delley},\ and\ \citenamefont {Greber}}]{wes15}%
	\BibitemOpen
	\bibfield  {author} {\bibinfo {author} {\bibfnamefont {R.}~\bibnamefont
			{Westerstr\"om}}, \bibinfo {author} {\bibfnamefont {A.-C.}\ \bibnamefont
			{Uldry}}, \bibinfo {author} {\bibfnamefont {R.}~\bibnamefont {Stania}},
		\bibinfo {author} {\bibfnamefont {J.}~\bibnamefont {Dreiser}}, \bibinfo
		{author} {\bibfnamefont {C.}~\bibnamefont {Piamonteze}}, \bibinfo {author}
		{\bibfnamefont {M.}~\bibnamefont {Muntwiler}}, \bibinfo {author}
		{\bibfnamefont {F.}~\bibnamefont {Matsui}}, \bibinfo {author} {\bibfnamefont
			{S.}~\bibnamefont {Rusponi}}, \bibinfo {author} {\bibfnamefont
			{H.}~\bibnamefont {Brune}}, \bibinfo {author} {\bibfnamefont
			{S.}~\bibnamefont {Yang}}, \bibinfo {author} {\bibfnamefont {A.}~\bibnamefont
			{Popov}}, \bibinfo {author} {\bibfnamefont {B.}~\bibnamefont {B\"uchner}},
		\bibinfo {author} {\bibfnamefont {B.}~\bibnamefont {Delley}}, \ and\ \bibinfo
		{author} {\bibfnamefont {T.}~\bibnamefont {Greber}},\ }\href {\doibase
		10.1103/PhysRevLett.114.087201} {\bibfield  {journal} {\bibinfo  {journal}
			{Phys. Rev. Lett.}\ }\textbf {\bibinfo {volume} {114}},\ \bibinfo {pages}
		{087201} (\bibinfo {year} {2015})}\BibitemShut {NoStop}%
	\bibitem [{\citenamefont {Heinrich}\ \emph {et~al.}(2004)\citenamefont
		{Heinrich}, \citenamefont {Gupta}, \citenamefont {Lutz},\ and\ \citenamefont
		{Eigler}}]{hei04}%
	\BibitemOpen
	\bibfield  {author} {\bibinfo {author} {\bibfnamefont {A.~J.}\ \bibnamefont
			{Heinrich}}, \bibinfo {author} {\bibfnamefont {J.~A.}\ \bibnamefont {Gupta}},
		\bibinfo {author} {\bibfnamefont {C.~P.}\ \bibnamefont {Lutz}}, \ and\
		\bibinfo {author} {\bibfnamefont {D.~M.}\ \bibnamefont {Eigler}},\ }\href
	{\doibase 10.1126/science.1101077} {\bibfield  {journal} {\bibinfo  {journal}
			{Science}\ }\textbf {\bibinfo {volume} {306}},\ \bibinfo {pages} {466}
		(\bibinfo {year} {2004})}\BibitemShut {NoStop}%
	\bibitem [{\citenamefont {Coffey}\ \emph {et~al.}(2015)\citenamefont {Coffey},
		\citenamefont {Diez-Ferrer}, \citenamefont {Serrate}, \citenamefont {Ciria},
		\citenamefont {Fuente},\ and\ \citenamefont {Arnaudas}}]{cof15}%
	\BibitemOpen
	\bibfield  {author} {\bibinfo {author} {\bibfnamefont {D.}~\bibnamefont
			{Coffey}}, \bibinfo {author} {\bibfnamefont {J.~L.}\ \bibnamefont
			{Diez-Ferrer}}, \bibinfo {author} {\bibfnamefont {D.}~\bibnamefont
			{Serrate}}, \bibinfo {author} {\bibfnamefont {M.}~\bibnamefont {Ciria}},
		\bibinfo {author} {\bibfnamefont {C.~d.~l.}\ \bibnamefont {Fuente}}, \ and\
		\bibinfo {author} {\bibfnamefont {J.~I.}\ \bibnamefont {Arnaudas}},\ }\href
	{\doibase 10.1038/srep13709} {\bibfield  {journal} {\bibinfo  {journal} {Sci.
				Rep.}\ }\textbf {\bibinfo {volume} {5}},\ \bibinfo {pages} {13709} (\bibinfo
		{year} {2015})}\BibitemShut {NoStop}%
	\bibitem [{\citenamefont {Steinbrecher}\ \emph {et~al.}(2015)\citenamefont
		{Steinbrecher}, \citenamefont {Sonntag}, \citenamefont {dos Santos~Dias},
		\citenamefont {Bouhassoune}, \citenamefont {Lounis}, \citenamefont {Wiebe},
		\citenamefont {Wiesendanger},\ and\ \citenamefont {Khajetoorians}}]{ste15}%
	\BibitemOpen
	\bibfield  {author} {\bibinfo {author} {\bibfnamefont {M.}~\bibnamefont
			{Steinbrecher}}, \bibinfo {author} {\bibfnamefont {A.}~\bibnamefont
			{Sonntag}}, \bibinfo {author} {\bibfnamefont {M.}~\bibnamefont {dos
				Santos~Dias}}, \bibinfo {author} {\bibfnamefont {M.}~\bibnamefont
			{Bouhassoune}}, \bibinfo {author} {\bibfnamefont {S.}~\bibnamefont {Lounis}},
		\bibinfo {author} {\bibfnamefont {J.}~\bibnamefont {Wiebe}}, \bibinfo
		{author} {\bibfnamefont {R.}~\bibnamefont {Wiesendanger}}, \ and\ \bibinfo
		{author} {\bibfnamefont {A.~A.}\ \bibnamefont {Khajetoorians}},\ }\href
	{\doibase 10.1038/ncomms10454} {\bibfield  {journal} {\bibinfo  {journal}
			{Nat. Commun.}\ }\textbf {\bibinfo {volume} {7}},\ \bibinfo {pages} {10454}
		(\bibinfo {year} {2015})}\BibitemShut {NoStop}%
	\bibitem [{\citenamefont {{Natterer}}\ \emph {et~al.}(2017)\citenamefont
		{{Natterer}}, \citenamefont {{Yang}}, \citenamefont {{Paul}}, \citenamefont
		{{Willke}}, \citenamefont {{Choi}}, \citenamefont {{Greber}}, \citenamefont
		{{Heinrich}},\ and\ \citenamefont {{Lutz}}}]{nat17}%
	\BibitemOpen
	\bibfield  {author} {\bibinfo {author} {\bibfnamefont {F.~D.}\ \bibnamefont
			{{Natterer}}}, \bibinfo {author} {\bibfnamefont {K.}~\bibnamefont {{Yang}}},
		\bibinfo {author} {\bibfnamefont {W.}~\bibnamefont {{Paul}}}, \bibinfo
		{author} {\bibfnamefont {P.}~\bibnamefont {{Willke}}}, \bibinfo {author}
		{\bibfnamefont {T.}~\bibnamefont {{Choi}}}, \bibinfo {author} {\bibfnamefont
			{T.}~\bibnamefont {{Greber}}}, \bibinfo {author} {\bibfnamefont {A.~J.}\
			\bibnamefont {{Heinrich}}}, \ and\ \bibinfo {author} {\bibfnamefont {C.~P.}\
			\bibnamefont {{Lutz}}},\ }\href {\doibase 10.1038/nature21371} {\bibfield
		{journal} {\bibinfo  {journal} {Nature}\ }\textbf {\bibinfo {volume} {543}},\
		\bibinfo {pages} {226} (\bibinfo {year} {2017})}\BibitemShut {NoStop}%
	\bibitem [{\citenamefont {Bode}\ \emph {et~al.}(1998)\citenamefont {Bode},
		\citenamefont {Getzlaff},\ and\ \citenamefont {Wiesendanger}}]{mat98}%
	\BibitemOpen
	\bibfield  {author} {\bibinfo {author} {\bibfnamefont {M.}~\bibnamefont
			{Bode}}, \bibinfo {author} {\bibfnamefont {M.}~\bibnamefont {Getzlaff}}, \
		and\ \bibinfo {author} {\bibfnamefont {R.}~\bibnamefont {Wiesendanger}},\
	}\href {\doibase 10.1103/PhysRevLett.81.4256} {\bibfield  {journal} {\bibinfo
			{journal} {Phys. Rev. Lett.}\ }\textbf {\bibinfo {volume} {81}},\ \bibinfo
		{pages} {4256} (\bibinfo {year} {1998})}\BibitemShut {NoStop}%
	\bibitem [{\citenamefont {Rau}\ \emph {et~al.}(2014)\citenamefont {Rau},
		\citenamefont {Baumann}, \citenamefont {Rusponi}, \citenamefont {Donati},
		\citenamefont {Stepanow}, \citenamefont {Gragnaniello}, \citenamefont
		{Dreiser}, \citenamefont {Piamonteze}, \citenamefont {Nolting}, \citenamefont
		{Gangopadhyay}, \citenamefont {Albertini}, \citenamefont {Macfarlane},
		\citenamefont {Lutz}, \citenamefont {Jones}, \citenamefont {Gambardella},
		\citenamefont {Heinrich},\ and\ \citenamefont {Brune}}]{rau14}%
	\BibitemOpen
	\bibfield  {author} {\bibinfo {author} {\bibfnamefont {I.~G.}\ \bibnamefont
			{Rau}}, \bibinfo {author} {\bibfnamefont {S.}~\bibnamefont {Baumann}},
		\bibinfo {author} {\bibfnamefont {S.}~\bibnamefont {Rusponi}}, \bibinfo
		{author} {\bibfnamefont {F.}~\bibnamefont {Donati}}, \bibinfo {author}
		{\bibfnamefont {S.}~\bibnamefont {Stepanow}}, \bibinfo {author}
		{\bibfnamefont {L.}~\bibnamefont {Gragnaniello}}, \bibinfo {author}
		{\bibfnamefont {J.}~\bibnamefont {Dreiser}}, \bibinfo {author} {\bibfnamefont
			{C.}~\bibnamefont {Piamonteze}}, \bibinfo {author} {\bibfnamefont
			{F.}~\bibnamefont {Nolting}}, \bibinfo {author} {\bibfnamefont
			{S.}~\bibnamefont {Gangopadhyay}}, \bibinfo {author} {\bibfnamefont {O.~R.}\
			\bibnamefont {Albertini}}, \bibinfo {author} {\bibfnamefont {R.~M.}\
			\bibnamefont {Macfarlane}}, \bibinfo {author} {\bibfnamefont {C.~P.}\
			\bibnamefont {Lutz}}, \bibinfo {author} {\bibfnamefont {B.~A.}\ \bibnamefont
			{Jones}}, \bibinfo {author} {\bibfnamefont {P.}~\bibnamefont {Gambardella}},
		\bibinfo {author} {\bibfnamefont {A.~J.}\ \bibnamefont {Heinrich}}, \ and\
		\bibinfo {author} {\bibfnamefont {H.}~\bibnamefont {Brune}},\ }\href
	{\doibase 10.1126/science.1252841} {\bibfield  {journal} {\bibinfo  {journal}
			{Science}\ }\textbf {\bibinfo {volume} {344}},\ \bibinfo {pages} {988}
		(\bibinfo {year} {2014})}\BibitemShut {NoStop}%
	\bibitem [{\citenamefont {Donati}\ \emph {et~al.}(2016)\citenamefont {Donati},
		\citenamefont {Rusponi}, \citenamefont {Stepanow}, \citenamefont
		{W\"{a}ckerlin}, \citenamefont {Singha}, \citenamefont {Persichetti},
		\citenamefont {Baltic}, \citenamefont {Diller}, \citenamefont {Patthey},
		\citenamefont {Fernandes}, \citenamefont {Dreiser}, \citenamefont {{\v
				S}ljivan{\v c}anin}, \citenamefont {Kummer}, \citenamefont {Nistor},
		\citenamefont {Gambardella},\ and\ \citenamefont {Brune}}]{don16}%
	\BibitemOpen
	\bibfield  {author} {\bibinfo {author} {\bibfnamefont {F.}~\bibnamefont
			{Donati}}, \bibinfo {author} {\bibfnamefont {S.}~\bibnamefont {Rusponi}},
		\bibinfo {author} {\bibfnamefont {S.}~\bibnamefont {Stepanow}}, \bibinfo
		{author} {\bibfnamefont {C.}~\bibnamefont {W\"{a}ckerlin}}, \bibinfo {author}
		{\bibfnamefont {A.}~\bibnamefont {Singha}}, \bibinfo {author} {\bibfnamefont
			{L.}~\bibnamefont {Persichetti}}, \bibinfo {author} {\bibfnamefont
			{R.}~\bibnamefont {Baltic}}, \bibinfo {author} {\bibfnamefont
			{K.}~\bibnamefont {Diller}}, \bibinfo {author} {\bibfnamefont
			{F.}~\bibnamefont {Patthey}}, \bibinfo {author} {\bibfnamefont
			{E.}~\bibnamefont {Fernandes}}, \bibinfo {author} {\bibfnamefont
			{J.}~\bibnamefont {Dreiser}}, \bibinfo {author} {\bibfnamefont {{\v
					Z}.}~\bibnamefont {{\v S}ljivan{\v c}anin}}, \bibinfo {author} {\bibfnamefont
			{K.}~\bibnamefont {Kummer}}, \bibinfo {author} {\bibfnamefont
			{C.}~\bibnamefont {Nistor}}, \bibinfo {author} {\bibfnamefont
			{P.}~\bibnamefont {Gambardella}}, \ and\ \bibinfo {author} {\bibfnamefont
			{H.}~\bibnamefont {Brune}},\ }\href {\doibase 10.1126/science.aad9898}
	{\bibfield  {journal} {\bibinfo  {journal} {Science}\ }\textbf {\bibinfo
			{volume} {352}},\ \bibinfo {pages} {318} (\bibinfo {year}
		{2016})}\BibitemShut {NoStop}%
	\bibitem [{\citenamefont {W{\"a}ckerlin}\ \emph {et~al.}(2016)\citenamefont
		{W{\"a}ckerlin}, \citenamefont {Donati}, \citenamefont {Singha},
		\citenamefont {Baltic}, \citenamefont {Rusponi}, \citenamefont {Diller},
		\citenamefont {Patthey}, \citenamefont {Pivetta}, \citenamefont {Lan},
		\citenamefont {Klyatskaya}, \citenamefont {Ruben}, \citenamefont {Brune},\
		and\ \citenamefont {Dreiser}}]{wac16}%
	\BibitemOpen
	\bibfield  {author} {\bibinfo {author} {\bibfnamefont {C.}~\bibnamefont
			{W{\"a}ckerlin}}, \bibinfo {author} {\bibfnamefont {F.}~\bibnamefont
			{Donati}}, \bibinfo {author} {\bibfnamefont {A.}~\bibnamefont {Singha}},
		\bibinfo {author} {\bibfnamefont {R.}~\bibnamefont {Baltic}}, \bibinfo
		{author} {\bibfnamefont {S.}~\bibnamefont {Rusponi}}, \bibinfo {author}
		{\bibfnamefont {K.}~\bibnamefont {Diller}}, \bibinfo {author} {\bibfnamefont
			{F.}~\bibnamefont {Patthey}}, \bibinfo {author} {\bibfnamefont
			{M.}~\bibnamefont {Pivetta}}, \bibinfo {author} {\bibfnamefont
			{Y.}~\bibnamefont {Lan}}, \bibinfo {author} {\bibfnamefont {S.}~\bibnamefont
			{Klyatskaya}}, \bibinfo {author} {\bibfnamefont {M.}~\bibnamefont {Ruben}},
		\bibinfo {author} {\bibfnamefont {H.}~\bibnamefont {Brune}}, \ and\ \bibinfo
		{author} {\bibfnamefont {J.}~\bibnamefont {Dreiser}},\ }\href {\doibase
		10.1002/adma.201506305} {\bibfield  {journal} {\bibinfo  {journal} {Adv.
				Mater.}\ }\textbf {\bibinfo {volume} {28}},\ \bibinfo {pages} {5195}
		(\bibinfo {year} {2016})}\BibitemShut {NoStop}%
	\bibitem [{\citenamefont {Fernandes}\ \emph {et~al.}(2017)\citenamefont
		{Fernandes}, \citenamefont {Donati}, \citenamefont {Patthey}, \citenamefont
		{Stavri\'{c}}, \citenamefont {{\v S}ljivan{\v c}anin},\ and\ \citenamefont
		{Brune}}]{fer17}%
	\BibitemOpen
	\bibfield  {author} {\bibinfo {author} {\bibfnamefont {E.}~\bibnamefont
			{Fernandes}}, \bibinfo {author} {\bibfnamefont {F.}~\bibnamefont {Donati}},
		\bibinfo {author} {\bibfnamefont {F.}~\bibnamefont {Patthey}}, \bibinfo
		{author} {\bibfnamefont {S.}~\bibnamefont {Stavri\'{c}}}, \bibinfo {author}
		{\bibfnamefont {{\v Z}.}~\bibnamefont {{\v S}ljivan{\v c}anin}}, \ and\
		\bibinfo {author} {\bibfnamefont {H.}~\bibnamefont {Brune}},\ }\href
	{\doibase 10.1103/PhysRevB.96.045419} {\bibfield  {journal} {\bibinfo
			{journal} {Phys. Rev. B}\ }\textbf {\bibinfo {volume} {96}},\ \bibinfo
		{pages} {045419} (\bibinfo {year} {2017})}\BibitemShut {NoStop}%
	\bibitem [{\citenamefont {Claude}(2005)}]{cla05}%
	\BibitemOpen
	\bibfield  {author} {\bibinfo {author} {\bibfnamefont {L.}~\bibnamefont
			{Claude}},\ }\emph {\bibinfo {title} {Construction d'un microscope \'a effet
			tunnel \'a basse temp\'erature et \'etudes d'impuret\'es magn\'etiques en
			surfaces}},\ \href {\doibase doi:10.5075/epfl-thesis-3276} {Ph.D. thesis},\
	\bibinfo  {school} {\'Ecole Polytechnique F\'ed\'eral de Lausanne} (\bibinfo
	{year} {2005})\BibitemShut {NoStop}%
	\bibitem [{\citenamefont {Baumann}\ \emph {et~al.}(2015)\citenamefont
		{Baumann}, \citenamefont {Donati}, \citenamefont {Stepanow}, \citenamefont
		{Rusponi}, \citenamefont {Paul}, \citenamefont {Gangopadhyay}, \citenamefont
		{Rau}, \citenamefont {Pacchioni}, \citenamefont {Gragnaniello}, \citenamefont
		{Pivetta}, \citenamefont {Dreiser}, \citenamefont {Piamonteze}, \citenamefont
		{Lutz}, \citenamefont {Macfarlane}, \citenamefont {Jones}, \citenamefont
		{Gambardella}, \citenamefont {Heinrich},\ and\ \citenamefont
		{Brune}}]{bau15}%
	\BibitemOpen
	\bibfield  {author} {\bibinfo {author} {\bibfnamefont {S.}~\bibnamefont
			{Baumann}}, \bibinfo {author} {\bibfnamefont {F.}~\bibnamefont {Donati}},
		\bibinfo {author} {\bibfnamefont {S.}~\bibnamefont {Stepanow}}, \bibinfo
		{author} {\bibfnamefont {S.}~\bibnamefont {Rusponi}}, \bibinfo {author}
		{\bibfnamefont {W.}~\bibnamefont {Paul}}, \bibinfo {author} {\bibfnamefont
			{S.}~\bibnamefont {Gangopadhyay}}, \bibinfo {author} {\bibfnamefont {I.~G.}\
			\bibnamefont {Rau}}, \bibinfo {author} {\bibfnamefont {G.~E.}\ \bibnamefont
			{Pacchioni}}, \bibinfo {author} {\bibfnamefont {L.}~\bibnamefont
			{Gragnaniello}}, \bibinfo {author} {\bibfnamefont {M.}~\bibnamefont
			{Pivetta}}, \bibinfo {author} {\bibfnamefont {J.}~\bibnamefont {Dreiser}},
		\bibinfo {author} {\bibfnamefont {C.}~\bibnamefont {Piamonteze}}, \bibinfo
		{author} {\bibfnamefont {C.~P.}\ \bibnamefont {Lutz}}, \bibinfo {author}
		{\bibfnamefont {R.~M.}\ \bibnamefont {Macfarlane}}, \bibinfo {author}
		{\bibfnamefont {B.~A.}\ \bibnamefont {Jones}}, \bibinfo {author}
		{\bibfnamefont {P.}~\bibnamefont {Gambardella}}, \bibinfo {author}
		{\bibfnamefont {A.~J.}\ \bibnamefont {Heinrich}}, \ and\ \bibinfo {author}
		{\bibfnamefont {H.}~\bibnamefont {Brune}},\ }\href {\doibase
		10.1103/PhysRevLett.115.237202} {\bibfield  {journal} {\bibinfo  {journal}
			{Phys. Rev. Lett.}\ }\textbf {\bibinfo {volume} {115}},\ \bibinfo {pages}
		{237202} (\bibinfo {year} {2015})}\BibitemShut {NoStop}%
	\bibitem [{\citenamefont {Paul}\ \emph {et~al.}(2016)\citenamefont {Paul},
		\citenamefont {Yang}, \citenamefont {Baumann}, \citenamefont {Romming},
		\citenamefont {Choi}, \citenamefont {Lutz},\ and\ \citenamefont
		{Heinrich}}]{pau16}%
	\BibitemOpen
	\bibfield  {author} {\bibinfo {author} {\bibfnamefont {W.}~\bibnamefont
			{Paul}}, \bibinfo {author} {\bibfnamefont {K.}~\bibnamefont {Yang}}, \bibinfo
		{author} {\bibfnamefont {S.}~\bibnamefont {Baumann}}, \bibinfo {author}
		{\bibfnamefont {N.}~\bibnamefont {Romming}}, \bibinfo {author} {\bibfnamefont
			{T.}~\bibnamefont {Choi}}, \bibinfo {author} {\bibfnamefont {C.~P.}\
			\bibnamefont {Lutz}}, \ and\ \bibinfo {author} {\bibfnamefont {A.~J.}\
			\bibnamefont {Heinrich}},\ }\href {\doibase 10.1038/nphys3965} {\bibfield
		{journal} {\bibinfo  {journal} {Nat. Phys.}\ }\textbf {\bibinfo {volume}
			{13}},\ \bibinfo {pages} {403} (\bibinfo {year} {2016})}\BibitemShut
	{NoStop}%
	\bibitem [{\citenamefont {Schuh}\ \emph {et~al.}(2010)\citenamefont {Schuh},
		\citenamefont {Balashov}, \citenamefont {Miyamachi}, \citenamefont
		{Tak\'acs}, \citenamefont {Suga},\ and\ \citenamefont {Wulfhekel}}]{sch10}%
	\BibitemOpen
	\bibfield  {author} {\bibinfo {author} {\bibfnamefont {T.}~\bibnamefont
			{Schuh}}, \bibinfo {author} {\bibfnamefont {T.}~\bibnamefont {Balashov}},
		\bibinfo {author} {\bibfnamefont {T.}~\bibnamefont {Miyamachi}}, \bibinfo
		{author} {\bibfnamefont {A.~F.}\ \bibnamefont {Tak\'acs}}, \bibinfo {author}
		{\bibfnamefont {S.}~\bibnamefont {Suga}}, \ and\ \bibinfo {author}
		{\bibfnamefont {W.}~\bibnamefont {Wulfhekel}},\ }\href {\doibase
		10.1063/1.3365113} {\bibfield  {journal} {\bibinfo  {journal} {J. Appl.
				Phys.}\ }\textbf {\bibinfo {volume} {107}},\ \bibinfo {pages} {09E156}
		(\bibinfo {year} {2010})}\BibitemShut {NoStop}%
	\bibitem [{\citenamefont {Baumann}(2015)}]{bau15b}%
	\BibitemOpen
	\bibfield  {author} {\bibinfo {author} {\bibfnamefont {S.}~\bibnamefont
			{Baumann}},\ }\emph {\bibinfo {title} {Investigation of the unusual magnetic
			properties of Fe and Co on MgO with high spatial, energy and temporal
			resolution}},\ \href {\doibase 10.5451/unibas-006489486} {Ph.D. thesis},\
	\bibinfo  {school} {University of Basel} (\bibinfo {year} {2015})\BibitemShut
	{NoStop}%
	\bibitem [{\citenamefont {Loth}\ \emph {et~al.}(2010)\citenamefont {Loth},
		\citenamefont {von Bergmann}, \citenamefont {Ternes}, \citenamefont {Otte},
		\citenamefont {Lutz},\ and\ \citenamefont {Heinrich}}]{lot10}%
	\BibitemOpen
	\bibfield  {author} {\bibinfo {author} {\bibfnamefont {S.}~\bibnamefont
			{Loth}}, \bibinfo {author} {\bibfnamefont {K.}~\bibnamefont {von Bergmann}},
		\bibinfo {author} {\bibfnamefont {M.}~\bibnamefont {Ternes}}, \bibinfo
		{author} {\bibfnamefont {A.~F.}\ \bibnamefont {Otte}}, \bibinfo {author}
		{\bibfnamefont {C.~P.}\ \bibnamefont {Lutz}}, \ and\ \bibinfo {author}
		{\bibfnamefont {A.~J.}\ \bibnamefont {Heinrich}},\ }\href
	{http://dx.doi.org/10.1038/nphys1616} {\bibfield  {journal} {\bibinfo
			{journal} {Nat. Phys.}\ }\textbf {\bibinfo {volume} {6}},\ \bibinfo {pages}
		{340} (\bibinfo {year} {2010})}\BibitemShut {NoStop}%
	\bibitem [{\citenamefont {Dreiser}(2015)}]{jan15}%
	\BibitemOpen
	\bibfield  {author} {\bibinfo {author} {\bibfnamefont {J.}~\bibnamefont
			{Dreiser}},\ }\href {http://stacks.iop.org/0953-8984/27/i=18/a=183203}
	{\bibfield  {journal} {\bibinfo  {journal} {J. Phys: Condensed Matter}\
		}\textbf {\bibinfo {volume} {27}},\ \bibinfo {pages} {183203} (\bibinfo
		{year} {2015})}\BibitemShut {NoStop}%
	\bibitem [{\citenamefont {Hirjibehedin}\ \emph {et~al.}(2007)\citenamefont
		{Hirjibehedin}, \citenamefont {Lin}, \citenamefont {Otte}, \citenamefont
		{Ternes}, \citenamefont {Lutz}, \citenamefont {Jones},\ and\ \citenamefont
		{Heinrich}}]{hir07}%
	\BibitemOpen
	\bibfield  {author} {\bibinfo {author} {\bibfnamefont {C.~F.}\ \bibnamefont
			{Hirjibehedin}}, \bibinfo {author} {\bibfnamefont {C.-Y.}\ \bibnamefont
			{Lin}}, \bibinfo {author} {\bibfnamefont {A.~F.}\ \bibnamefont {Otte}},
		\bibinfo {author} {\bibfnamefont {M.}~\bibnamefont {Ternes}}, \bibinfo
		{author} {\bibfnamefont {C.~P.}\ \bibnamefont {Lutz}}, \bibinfo {author}
		{\bibfnamefont {B.~A.}\ \bibnamefont {Jones}}, \ and\ \bibinfo {author}
		{\bibfnamefont {A.~J.}\ \bibnamefont {Heinrich}},\ }\href {\doibase
		10.1126/science.1146110} {\bibfield  {journal} {\bibinfo  {journal}
			{Science}\ }\textbf {\bibinfo {volume} {317}},\ \bibinfo {pages} {1199}
		(\bibinfo {year} {2007})}\BibitemShut {NoStop}%
	\bibitem [{\citenamefont {Otte}\ \emph {et~al.}(2008)\citenamefont {Otte},
		\citenamefont {Ternes}, \citenamefont {von Bergmann}, \citenamefont {Loth},
		\citenamefont {Brune}, \citenamefont {Lutz}, \citenamefont {Hirjibehedin},\
		and\ \citenamefont {Heinrich}}]{ott08}%
	\BibitemOpen
	\bibfield  {author} {\bibinfo {author} {\bibfnamefont {A.~F.}\ \bibnamefont
			{Otte}}, \bibinfo {author} {\bibfnamefont {M.}~\bibnamefont {Ternes}},
		\bibinfo {author} {\bibfnamefont {K.}~\bibnamefont {von Bergmann}}, \bibinfo
		{author} {\bibfnamefont {S.}~\bibnamefont {Loth}}, \bibinfo {author}
		{\bibfnamefont {H.}~\bibnamefont {Brune}}, \bibinfo {author} {\bibfnamefont
			{C.~P.}\ \bibnamefont {Lutz}}, \bibinfo {author} {\bibfnamefont {C.~F.}\
			\bibnamefont {Hirjibehedin}}, \ and\ \bibinfo {author} {\bibfnamefont
			{A.~J.}\ \bibnamefont {Heinrich}},\ }\href
	{https://www.nature.com/articles/nphys1072} {\bibfield  {journal} {\bibinfo
			{journal} {Nat. Phys.}\ }\textbf {\bibinfo {volume} {4}},\ \bibinfo {pages}
		{847} (\bibinfo {year} {2008})}\BibitemShut {NoStop}%
	\bibitem [{\citenamefont {Janssen}(2003)}]{jan03}%
	\BibitemOpen
	\bibfield  {author} {\bibinfo {author} {\bibfnamefont {Y.}~\bibnamefont
			{Janssen}},\ }\emph {\bibinfo {title} {Interplay between magnetic anisotropy
			and exchange interactions in rare-earth-transition-metal ferrimagnets}},\
	\href {http://hdl.handle.net/11245/1.202364} {Ph.D. thesis},\ \bibinfo
	{school} {Universiteit van Amsterdam} (\bibinfo {year} {2003})\BibitemShut
	{NoStop}%
	\bibitem [{\citenamefont {Gambardella}\ \emph {et~al.}(2003)\citenamefont
		{Gambardella}, \citenamefont {Rusponi}, \citenamefont {Veronese},
		\citenamefont {Dhesi}, \citenamefont {Grazioli}, \citenamefont {Dallmeyer},
		\citenamefont {Cabria}, \citenamefont {Zeller}, \citenamefont {Dederichs},
		\citenamefont {Kern}, \citenamefont {Carbone},\ and\ \citenamefont
		{Brune}}]{gam03}%
	\BibitemOpen
	\bibfield  {author} {\bibinfo {author} {\bibfnamefont {P.}~\bibnamefont
			{Gambardella}}, \bibinfo {author} {\bibfnamefont {S.}~\bibnamefont
			{Rusponi}}, \bibinfo {author} {\bibfnamefont {M.}~\bibnamefont {Veronese}},
		\bibinfo {author} {\bibfnamefont {S.~S.}\ \bibnamefont {Dhesi}}, \bibinfo
		{author} {\bibfnamefont {C.}~\bibnamefont {Grazioli}}, \bibinfo {author}
		{\bibfnamefont {A.}~\bibnamefont {Dallmeyer}}, \bibinfo {author}
		{\bibfnamefont {I.}~\bibnamefont {Cabria}}, \bibinfo {author} {\bibfnamefont
			{R.}~\bibnamefont {Zeller}}, \bibinfo {author} {\bibfnamefont {P.~H.}\
			\bibnamefont {Dederichs}}, \bibinfo {author} {\bibfnamefont {K.}~\bibnamefont
			{Kern}}, \bibinfo {author} {\bibfnamefont {C.}~\bibnamefont {Carbone}}, \
		and\ \bibinfo {author} {\bibfnamefont {H.}~\bibnamefont {Brune}},\ }\href
	{\doibase 10.1126/science.1082857} {\bibfield  {journal} {\bibinfo  {journal}
			{Science}\ }\textbf {\bibinfo {volume} {300}},\ \bibinfo {pages} {1130}
		(\bibinfo {year} {2003})}\BibitemShut {NoStop}%
	\bibitem [{\citenamefont {Donati}\ \emph {et~al.}(2014)\citenamefont {Donati},
		\citenamefont {Gragnaniello}, \citenamefont {Cavallin}, \citenamefont
		{Natterer}, \citenamefont {Dubout}, \citenamefont {Pivetta}, \citenamefont
		{Patthey}, \citenamefont {Dreiser}, \citenamefont {Piamonteze}, \citenamefont
		{Rusponi},\ and\ \citenamefont {Brune}}]{don14}%
	\BibitemOpen
	\bibfield  {author} {\bibinfo {author} {\bibfnamefont {F.}~\bibnamefont
			{Donati}}, \bibinfo {author} {\bibfnamefont {L.}~\bibnamefont
			{Gragnaniello}}, \bibinfo {author} {\bibfnamefont {A.}~\bibnamefont
			{Cavallin}}, \bibinfo {author} {\bibfnamefont {F.~D.}\ \bibnamefont
			{Natterer}}, \bibinfo {author} {\bibfnamefont {Q.}~\bibnamefont {Dubout}},
		\bibinfo {author} {\bibfnamefont {M.}~\bibnamefont {Pivetta}}, \bibinfo
		{author} {\bibfnamefont {F.}~\bibnamefont {Patthey}}, \bibinfo {author}
		{\bibfnamefont {J.}~\bibnamefont {Dreiser}}, \bibinfo {author} {\bibfnamefont
			{C.}~\bibnamefont {Piamonteze}}, \bibinfo {author} {\bibfnamefont
			{S.}~\bibnamefont {Rusponi}}, \ and\ \bibinfo {author} {\bibfnamefont
			{H.}~\bibnamefont {Brune}},\ }\href {\doibase 10.1103/PhysRevLett.113.177201}
	{\bibfield  {journal} {\bibinfo  {journal} {Phys. Rev. Lett.}\ }\textbf
		{\bibinfo {volume} {113}},\ \bibinfo {pages} {177201} (\bibinfo {year}
		{2014})}\BibitemShut {NoStop}%
	\bibitem [{\citenamefont {Bryant}\ \emph {et~al.}(2013)\citenamefont {Bryant},
		\citenamefont {Spinelli}, \citenamefont {Wagenaar}, \citenamefont {Gerrits},\
		and\ \citenamefont {Otte}}]{bry13}%
	\BibitemOpen
	\bibfield  {author} {\bibinfo {author} {\bibfnamefont {B.}~\bibnamefont
			{Bryant}}, \bibinfo {author} {\bibfnamefont {A.}~\bibnamefont {Spinelli}},
		\bibinfo {author} {\bibfnamefont {J.~J.~T.}\ \bibnamefont {Wagenaar}},
		\bibinfo {author} {\bibfnamefont {M.}~\bibnamefont {Gerrits}}, \ and\
		\bibinfo {author} {\bibfnamefont {A.~F.}\ \bibnamefont {Otte}},\ }\href
	{\doibase 10.1103/PhysRevLett.111.127203} {\bibfield  {journal} {\bibinfo
			{journal} {Phys. Rev. Lett.}\ }\textbf {\bibinfo {volume} {111}},\ \bibinfo
		{pages} {127203} (\bibinfo {year} {2013})}\BibitemShut {NoStop}%
	\bibitem [{\citenamefont {Gatteschi}\ \emph {et~al.}(2006)\citenamefont
		{Gatteschi}, \citenamefont {Sessoli},\ and\ \citenamefont {Villain}}]{gat06}%
	\BibitemOpen
	\bibfield  {author} {\bibinfo {author} {\bibfnamefont {D.}~\bibnamefont
			{Gatteschi}}, \bibinfo {author} {\bibfnamefont {R.}~\bibnamefont {Sessoli}},
		\ and\ \bibinfo {author} {\bibfnamefont {J.}~\bibnamefont {Villain}},\ }in\
	\href {\doibase 10.1093/acprof:oso/9780198567530.001.0001} {\emph {\bibinfo
			{booktitle} {Molecular Nanomagnets}}}\ (\bibinfo  {publisher} {Oxford
		Scholarship Online},\ \bibinfo {year} {2006})\BibitemShut {NoStop}%
	\bibitem [{\citenamefont {Wernsdorfer}\ and\ \citenamefont
		{Sessoli}(1999)}]{wer99}%
	\BibitemOpen
	\bibfield  {author} {\bibinfo {author} {\bibfnamefont {W.}~\bibnamefont
			{Wernsdorfer}}\ and\ \bibinfo {author} {\bibfnamefont {R.}~\bibnamefont
			{Sessoli}},\ }\href {\doibase 10.1126/science.284.5411.133} {\bibfield
		{journal} {\bibinfo  {journal} {Science}\ }\textbf {\bibinfo {volume}
			{284}},\ \bibinfo {pages} {133} (\bibinfo {year} {1999})}\BibitemShut
	{NoStop}%
	\bibitem [{\citenamefont {Blum}(2011)}]{blu11}%
	\BibitemOpen
	\bibfield  {author} {\bibinfo {author} {\bibfnamefont {K.}~\bibnamefont
			{Blum}},\ }\href {\doibase 10.1007/978-1-4757-4931-1} {\emph {\bibinfo
			{title} {Density Matrix Theory and Applications}}}\ (\bibinfo  {publisher}
	{Springer},\ \bibinfo {year} {2011})\BibitemShut {NoStop}%
	\bibitem [{\citenamefont {Singha}(2017)}]{apa17}%
	\BibitemOpen
	\bibfield  {author} {\bibinfo {author} {\bibfnamefont {A.}~\bibnamefont
			{Singha}},\ }\emph {\bibinfo {title} {Magnetic properties of surface-adsorbed
			single rare earth atoms, molecules, and atomic scale clusters}},\ \href
	{https://infoscience.epfl.ch/record/232433?ln=en} {Ph.D. thesis},\ \bibinfo
	{school} {Ecole Polytechnique Federal de Lausanne} (\bibinfo {year}
	{2017})\BibitemShut {NoStop}%
	\bibitem [{\citenamefont {Ozaki}(2003)}]{oza03}%
	\BibitemOpen
	\bibfield  {author} {\bibinfo {author} {\bibfnamefont {T.}~\bibnamefont
			{Ozaki}},\ }\href {\doibase 10.1103/PhysRevB.67.155108} {\bibfield  {journal}
		{\bibinfo  {journal} {Phys. Rev. B}\ }\textbf {\bibinfo {volume} {67}},\
		\bibinfo {pages} {155108} (\bibinfo {year} {2003})}\BibitemShut {NoStop}%
	\bibitem [{\citenamefont {Morrison}\ \emph {et~al.}(1993)\citenamefont
		{Morrison}, \citenamefont {Bylander},\ and\ \citenamefont
		{Kleinman}}]{mor93}%
	\BibitemOpen
	\bibfield  {author} {\bibinfo {author} {\bibfnamefont {I.}~\bibnamefont
			{Morrison}}, \bibinfo {author} {\bibfnamefont {D.~M.}\ \bibnamefont
			{Bylander}}, \ and\ \bibinfo {author} {\bibfnamefont {L.}~\bibnamefont
			{Kleinman}},\ }\href {\doibase 10.1103/PhysRevB.47.6728} {\bibfield
		{journal} {\bibinfo  {journal} {Phys. Rev. B}\ }\textbf {\bibinfo {volume}
			{47}},\ \bibinfo {pages} {6728} (\bibinfo {year} {1993})}\BibitemShut
	{NoStop}%
	\bibitem [{ope()}]{openmx-site}%
	\BibitemOpen
	\href@noop {} {\enquote {\bibinfo {title} {Welcome to openmx, open source
				package for material explorer},}\ }\bibinfo {howpublished}
	{\url{http://www.openmx-square.org/}},\ \bibinfo {note} {accessed:
		2017-12-22}\BibitemShut {NoStop}%
	\bibitem [{\citenamefont {Perdew}\ \emph {et~al.}(1996)\citenamefont {Perdew},
		\citenamefont {Burke},\ and\ \citenamefont {Ernzerhof}}]{per96}%
	\BibitemOpen
	\bibfield  {author} {\bibinfo {author} {\bibfnamefont {J.~P.}\ \bibnamefont
			{Perdew}}, \bibinfo {author} {\bibfnamefont {K.}~\bibnamefont {Burke}}, \
		and\ \bibinfo {author} {\bibfnamefont {M.}~\bibnamefont {Ernzerhof}},\ }\href
	{\doibase 10.1103/PhysRevLett.77.3865} {\bibfield  {journal} {\bibinfo
			{journal} {Phys. Rev. Lett.}\ }\textbf {\bibinfo {volume} {77}},\ \bibinfo
		{pages} {3865} (\bibinfo {year} {1996})}\BibitemShut {NoStop}%
	\bibitem [{\citenamefont {Anisimov}\ \emph {et~al.}(1991)\citenamefont
		{Anisimov}, \citenamefont {Zaanen},\ and\ \citenamefont {Andersen}}]{ani91}%
	\BibitemOpen
	\bibfield  {author} {\bibinfo {author} {\bibfnamefont {V.~I.}\ \bibnamefont
			{Anisimov}}, \bibinfo {author} {\bibfnamefont {J.}~\bibnamefont {Zaanen}}, \
		and\ \bibinfo {author} {\bibfnamefont {O.~K.}\ \bibnamefont {Andersen}},\
	}\href {\doibase 10.1103/PhysRevB.44.943} {\bibfield  {journal} {\bibinfo
			{journal} {Phys. Rev. B}\ }\textbf {\bibinfo {volume} {44}},\ \bibinfo
		{pages} {943} (\bibinfo {year} {1991})}\BibitemShut {NoStop}%
	\bibitem [{\citenamefont {Monkhorst}\ and\ \citenamefont {Pack}(1976)}]{mon76}%
	\BibitemOpen
	\bibfield  {author} {\bibinfo {author} {\bibfnamefont {H.~J.}\ \bibnamefont
			{Monkhorst}}\ and\ \bibinfo {author} {\bibfnamefont {J.~D.}\ \bibnamefont
			{Pack}},\ }\href {\doibase 10.1103/PhysRevB.13.5188} {\bibfield  {journal}
		{\bibinfo  {journal} {Phys. Rev. B}\ }\textbf {\bibinfo {volume} {13}},\
		\bibinfo {pages} {5188} (\bibinfo {year} {1976})}\BibitemShut {NoStop}%
	\bibitem [{\citenamefont {Liu}\ and\ \citenamefont {Nocedal}(1989)}]{liu89}%
	\BibitemOpen
	\bibfield  {author} {\bibinfo {author} {\bibfnamefont {D.~C.}\ \bibnamefont
			{Liu}}\ and\ \bibinfo {author} {\bibfnamefont {J.}~\bibnamefont {Nocedal}},\
	}\href {\doibase 10.1007/BF01589116} {\bibfield  {journal} {\bibinfo
			{journal} {Math. Program.}\ }\textbf {\bibinfo {volume} {45}},\ \bibinfo
		{pages} {503} (\bibinfo {year} {1989})}\BibitemShut {NoStop}%
\end{thebibliography}

\widetext
\newpage
\begin{center}
\textbf{\large }
\end{center}
\setcounter{equation}{0}
\setcounter{table}{0}
\counterwithin{figure}{section}
\setcounter{figure}{0}
\makeatletter
\renewcommand{\theequation}{S\arabic{equation}}
\renewcommand{\thefigure}{S\arabic{figure}}
\renewcommand{\bibnumfmt}[1]{[#1]}
\renewcommand{\citenumfont}[1]{#1}
\section{Sample preparation}
Rare earth elements need special care as they are highly reactive and prone to oxidation. As a first step, a high purity (99.9~\%) rod of Ho was cleaned by filing off the oxidized surface layer until a shiny metal surface became visible. In order to minimize its exposure to ambient conditions, the Ho rod was immediately placed into one of the cells of a triple electron-beam evaporator. A similar strategy was followed with a high purity Co rod. After installing the rods in the triple evaporator, it was baked at $150^{\circ}$C for 48~hours. To further ensure purity of our samples, all rods were degassed for several days by operating the evaporator at parameters very close to the ones for actual deposition. The degassing was terminated when no further change in the base pressure was observed after switching the e-beam evaporator on or off.

The Ag(100) single crystal was prepared using several Ar$^{+}$ ion sputtering and annealing cycles ($T = 773$~K, $p_{\rm max} = 1 \times 10^{-9}$~mbar). MgO was grown by evaporating Mg from a Knudsen cell onto a clean Ag(100) substrate in a background oxygen pressure of $1 \times 10^{-6}$~mbar. Prior to this preparation, the Mg source was thoroughly degassed. During MgO growth, the Ag(100) crystal was kept at 773~K. After the Mg evaporation, the Ag(100) surface was allowed to slowly cool down to room temperature at a rate of 22~K/min. The temperature of Ag(100) during deposition and the speed of its post-deposition cool down, determine the thickness and morphology of the MgO layers~\cite{bau15, pau16}. The labelling of MgO thickness follows our previous work~\cite{fer17}. The transfer of samples from the preparation to the STM chamber was {\it in-situ} in UHV. The atomic species were evaporated onto the sample held in the STM. This implies lowering the STM from the center of the magnet to the sample transfer position and opening the thermal shields~\cite{cla05} which leads to a sample temperature between 13 and 15~K during deposition. At these temperatures both adsorbed atoms are immobile. The amount of Co and Ho deposited (0.015~ML) was chosen to optimize the yield of dimers, while keeping the species still sufficiently far apart to avoid any kind of interaction between them. 

\section{Spectroscopy of homodimers}
\begin{figure}[h!]
\begin{center}
\includegraphics{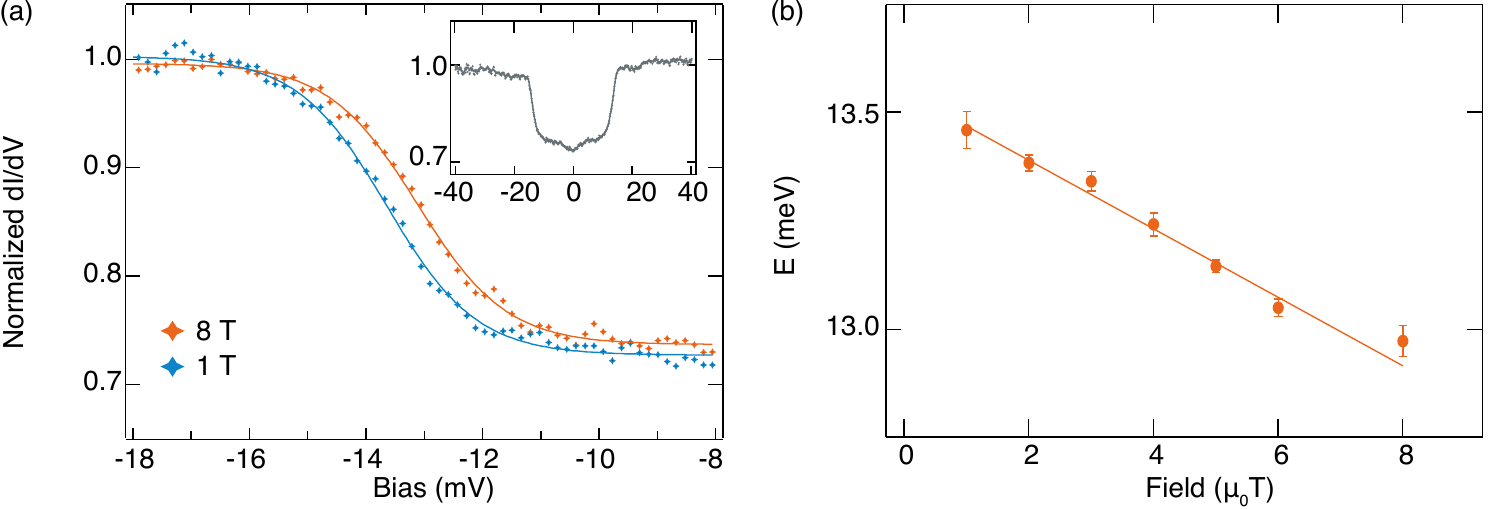}
\end{center}
\caption{(a) Zoom of the magnetic field dependent $dI/dV$ spectra of a Co dimer (dots: measurements, lines: sigmoid fits). Inset shows spectrum in the full energy range ($V_{\rm t} = 40$~mV, $I_{\rm t} = 250$~pA, $V_{\rm mod, ptp} = 1$~mV, $T = 4.3$~K). (b) Zeeman series of excitation energies, indicating the linear shift of the inelastic feature with the external magnetic field.}
\label{figS1}
\end{figure}

As mentioned in the main text, homodimers of Co and Ho, {\it i.e.}, Co$_2$ and Ho$_2$, exhibit their characteristic inelastic excitations. The $dI/dV$ steps of Co$_2$ and their Zeeman shift are shown in Fig.~\ref{figS1}. 
The inset of Fig.~\ref{figS1}(a) shows the entire spectrum, while the main figure shows the step for negative polarity and its 0.5~meV shift caused by an increase of the external out-of-plane magnetic field from 1 to 8~T. As expected, the shift is linear in field (Fig.~\ref{figS1}(b)). However, it is opposite in sign compared to the one of the $\pm 20$~meV step of HoCo, namely a field increase lowers the absolute value of the step energy. The reason is either a different orientation of the easy magnetization axis of the total dimer moment, or a different exchange coupling between the two atoms. 

\begin{figure}[h!]
\begin{center}
\includegraphics{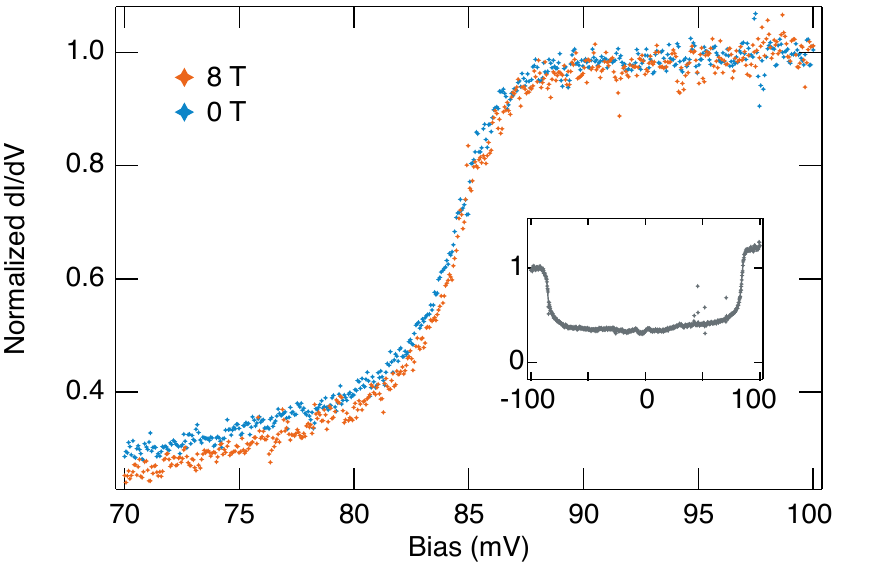}
\end{center}
\caption{Zoom into the positive bias $dI/dV$-step of a Ho dimer. The inset shows the $dI/dV$ spectrum in the full energy range. The step height is 50~\% \textit{i.e.}, the highest amongst the three investigated dimers. Also the excitation energy ($\pm 85$~meV) is by far the highest ($V_{\rm t} = 120$~mV, $I_{\rm t} = 1$~nA, $V_{\rm mod, ptp} = 200~\mu$V, $T = 0.4$~K).}
\label{figS2}
\end{figure}

Figure~\ref{figS2} shows the differential conductance steps observed on Ho$_2$. They are by far the most intense and highest energy steps amongst the three investigated dimer species. In addition, the Ho$_2$ $dI/dV$ steps distinguish themselves from the ones recorded on Co$_2$ and HoCo by the fact that they don't move in an external out-of-plane magnetic field, at least within our detection limits. This indicates either a non-magnetic origin or very strong in-plane magnetic anisotropy in these dimers. 

\section{Sigmoid fits to the conductance steps}
In order to determine most precisely the position of the inelastic differential conductance steps, we fitted $dI/dV$ with a sigmoid function using the Fermi-Dirac distribution $F(x)$ to generate the step: 
\begin{equation}
\begin{split}
f(x) & = A + M (1 - F(x)) \\
& = A + M \Big{(}1- \frac{1}{1 + e^\frac{x - E}{\delta}} \Big{)}
\end{split}
\label{eqn:1}
\end{equation}
$A$ and $M$ are the baseline and amplitude of the inelastic step, respectively, $x = {\rm e} V_{\rm t}$ is the energy corresponding to the tunnel voltage, and $E$ the energy of the inelastic step. The broadening of the step $\delta$ is caused by (a) the lifetime of the excited state involved in the inelastic process, (b) the temperature, and (c) the modulation voltage applied for recording the spectra with Lock-In detection. The fit parameters for the HoCo data shown in Fig.~\ref{SES}a and Fig.~\ref{SP-tip}b are listed in Table~\ref{tabS1}. Note that the tabulated $E$ values do not change within our precision if we analyze the numerically derived $d^2I/dV^2$ data and use a Gaussian profile for fitting it.

\begin{table}[h!]
\centering
\small
\begin{tabular}{c|r|r|r|r}
\hline
\hline
Parameters & \multicolumn{2}{c|}{Fig.~\ref{SES}(a)} & \multicolumn{2}{c}{Fig.~\ref{SP-tip}(b)}\\\cline{2-5} & 1~T & 8~T & $V_{\rm t} < 0$ & $V_{\rm t} > 0$ \\
\hline
\hline
$E$ (meV)	&	19.50&	20.80	&	8.30	&	7.90	\\	
\hline
$\delta$ (meV)	&	0.82	&	0.83	&	0.53	&	0.49	\\
\hline
\hline
\end{tabular}
\caption{Parameters obtained from the sigmoid fits of HoCo $dI/dV$ spectra.}
\label{tabS1}
\end{table}

Figure~\ref{SES}(b) shows in total 7 different HoCo species, 4 on 1 and 3 on 2~ML thick MgO(100) films, each of them represented with different colors. Each symbol results from sigmoid fits of the steps in the respective $dI/dV$-data. In Table~\ref{tabS2} we give the individual slopes of $E(\mu_0 H)$ that correspond to the effective $g$ factors, as well as the mean value for each MgO(100) thickness. 

\begin{table}[h!]
\centering
\small
\begin{tabular}{c|c|c|c}
\hline
\hline
MgO thickness & Species & effective $g$ factor & average effective $g$ factor\\
(ML) & &  & \\
\hline
\hline
& HoCo1 & $2.9 \pm 0.3$ & \\
& HoCo2 & $2.9 \pm0 .2$ & \\
$1$ & HoCo3 & $3.4 \pm 0.3$ & $3.1 \pm 0.3$\\
& HoCo4 & $3.2 \pm 0.2$ & \\
\hline
& HoCo5 & $2.4 \pm 0.4$ & \\
$2$	& HoCo6 & $2.5 \pm 0.4$ & $2.5 \pm 0.6$ \\
& HoCo7 & $2.8 \pm 0.9$ & \\		
\hline
\hline
\end{tabular}
\caption{Effective $g$ factors measured for the different heterodimers adsorbed on 1 and 2~ML of MgO shown with differently colored symbols in Fig.~\ref{SES}(b). They were obtained by linear regression to the data and are presented in the 3$^{\rm rd}$ column with errors representing $\sigma_{i}$ of the fit. The last column shows the mean value of $g$ for each layer thickness together with the corresponding standard deviation calculated as $\sqrt{\sum_{i}^{n} (\sigma_{i}^2/n)}$, where $n$ denotes the total number of cases.}
\label{tabS2}
\end{table}

\section{Spin Hamiltonian model}
The effective spin Hamiltonian (SH) model is frequently adopted for describing a system consisting of many spins. In this model, the individual contributions of orbital and spin moments are replaced by an effective spin moment $S$ that obeys the same symmetry properties. This approach has been widely used for interpreting inelastic spectroscopy~\cite{don14, hir07, bry13} and is also frequently used in molecular magnets~\cite{gat06, wer99}. A typical SH takes the following form:
\begin{equation}
\hat H = \hat H_{CF} + \hat H_{Zeeman}
\label{eqn:S1}
\end{equation}
$\hat H_{CF}=D\hat S_{z}^{2}$ defines the CF along $z$ axis which causes the splitting of the magnetic levels differing in $S_{z}$. Here $D$ is the uniaxial anisotropy term and the $z$ component of the spin operator $\bf \hat S$ is defined as $\hat S_{\rm z}$. The Zeeman term of the spin Hamiltonian, describes the interaction of the effective spin with the external magnetic field $\mu_{0} \bf H$ and this can be written as
\begin{equation}
\hat H_{Zeeman} = g\mu_{\rm B}{\bf \hat S}\cdot \mu_{0} \bf H
\label{eqn:S2}
\end{equation}
The effective $g$ factor connects the magnetic field and the effective spin vector, and $\mu_{\rm B}$ is the Bohr magneton.

Now we turn to the specific case of a heterodimer. In addition to the terms already introduced in equation~\ref{eqn:S1}, we consider an exchange term, $\hat H_{Exchange}$, for defining the interaction between the two effective spins of magnitude $S_{\rm Ho}$ and $S_{\rm Co}$. In order to describe the coupled system including all the relative orientations of the two individual spins, we are going to employ a density matrix formalism~\cite{blu11}. Within this formalism, and according to the Heisenberg coupling scheme, $\hat H_{Exchange}$ can be expressed as:
\begin{equation}
\hat H_{Exchange} = J {\bf \hat S}_{\rm Ho}\cdot {\bf \hat S}_{\rm Co}
\label{eqn:S3}
\end{equation}
where $J$ is the coupling constant and ${\bf \hat S}_{\rm Ho}$ and ${\bf \hat S}_{\rm Co}$ are the effective spin operators defined using the density matrix formalism. Altogether these reduce the effective SH for a system of two coupled spins subject to an out-of-plane magnetic field $\mu_{0}\bf \hat H$ as:
\begin{equation}
\begin{split}
\hat H & = D\hat S_{z}^{2}+ J ({\bf \hat S}_{\rm Ho}\cdot{\bf \hat S}_{\rm Co} ) +\mu_{\rm B} [g_{\rm Ho} {\rm \hat S}_{z \rm Ho}+ g_{\rm Co} {\rm \hat S}_{z \rm Co}] \cdot \mu_{0}\bf \hat H
\end{split}
\label{eqn:S8}
\end{equation}

We first diagonalize the SH to obtain the eigen-values and eigen-vectors. The expectation value of an operator is given by the tracing the product of the density matrix of the system with the operator itself~\cite{blu11, apa17}. Following this, we compute the expectation values of out-of-plane magnetic moments $\langle {\bf \hat S}_{z \rm Ho}\rangle$, $\langle {\bf \hat S}_{z \rm Co}\rangle$, and $\langle {\bf \hat S}_{z}\rangle$. They are the expected $z$-projected moments of Ho and Co atoms, and the $z$-projected overall moment of the heterodimer respectively. Finally, the energy level distribution of the magnetic levels shown Fig.~\ref{levels} is produced by plotting the eigen-values of the SH as a function of the respective $\langle {\bf \hat S}_{z}\rangle$ moments.

In order to avoid overparametrization, we consider only first order uniaxial magnetic anisotropy. Note the large $g$ factor as well as the position of the inelastic steps can also be reproduced with a non-vanishing first order off-diagonal term. However, SES is only sensitive to $\Delta S_{z} = \pm 1, 0$ transitions. Therefore with only two experimentally observed inelastic steps, we can reliably reproduce the lowest part of the full multiplet structure only, without being sensitive to its overall shape. As the in-plane term largely governs the mixing of the states and therefore influences the overall shape of the full multiplet structure, we can not comment with large confidence on the corresponding value.

\section{DFT calculations}
The non-collinear spin-polarized DFT calculations of the HoCo dimers on 2~ML thin MgO(100) films adsorbed pseudomorphically on Ag(100) were performed with the openMX computer package~\cite{oza03}. We used fully relativistic norm-conserving pseudopotentials~\cite{mor93} to describe the interaction of the ions with valence electrons. The Kohn-Sham wavefunctions were expanded within a basis set of the optimized pseudoatomic orbitals~\cite{openmx-site}. The exchange-correlation effects were described using the Perdew-Burke-Ernzerhof (PBE) functional~\cite{per96}. For Ho $4f$ states we added on-site Coulomb corrections~\cite{ani91} with $U = 5$~eV, as also used in our previous study of single Ho atoms on the same surface~\cite{don16}. The MgO(100)/Ag(100) surface was modeled with a ($3 \times 3$) unit cell containing nine Mg and nine O atoms per MgO(100) layer. The underlying Ag(100) substrate was represented by a three-layer slab with nine Ag atoms per fcc(100) layer. The supercell calculations were carried out with the theoretically optimized Ag lattice constants of 4.14~{\AA} and a Monkhorst-Pack mesh of 16~{\bf k}-points was used for sampling of the surface Brillouin zone~\cite{mon76}. Periodically repeated slabs were decoupled by at least 12~{\AA} thick vacuum. Numerical stability of the calculations was increased by populating electronic states according to the Fermi-Dirac distribution at $T = 300$~K. Atomic positions were relaxed utilizing the Broyden-Fletcher-Goldfarb-Shanno (BFGS) algorithm~\cite{liu89}.

\begin{figure}[h]
\centering
\includegraphics[width = 0.4 \linewidth]{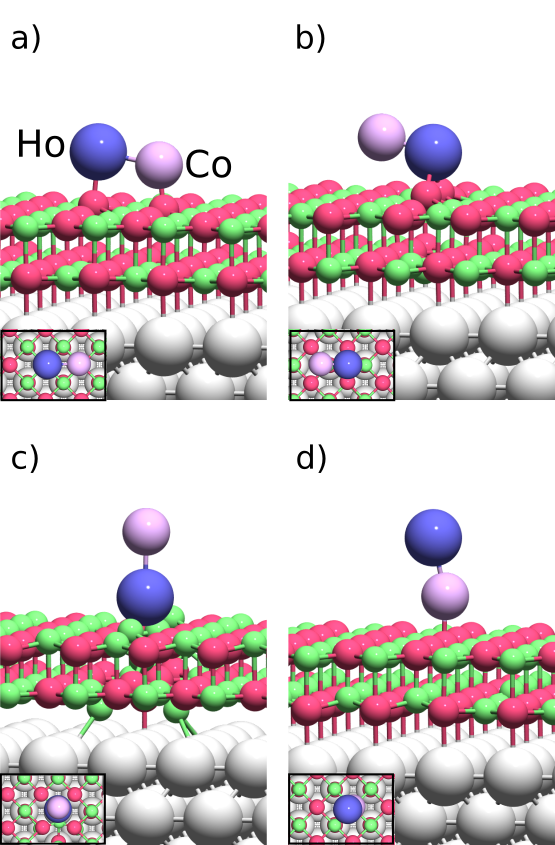}
\caption{Side views of adsorption geometries of HoCo dimer at 2ML-MgO(100)/Ag(100) considered in our DFT calculations. Insets show top views of the corresponding geometries.}
\label{figS3}
\end{figure}

We considered several adsorption geometries of the HoCo dimer on 2~ML MgO(100), as depicted in Fig.~\ref{figS3}. Based on the calculated total energies the structure shown in Figs.~1(d) and~\ref{figS3}(a) is identified as the most favorable one. The energies of other adsorption geometries with respect to this structure are listed in Table~\ref{tabS3}. Spin and orbital magnetic moments of Ho and Co as obtained from DFT calculations are also tabulated. For the most stable dimer structure, the Ho and Co magnetic moments are ferromagnetically coupled, their magnetic moments are nearly collinear, and point along the axis of the dimer. The first two properties are in agreement with experiment, while the orientation of the overall moment is in experiment close to the out-of-plane direction, implying that it is almost perpendicular to the dimer axis, in contrast to the DFT result. 

\begin{table*}
\centering
\begin{tabular}{c|cccccccc} 
\hline
adsorption geometry &\multicolumn{2}{c}{Fig. \ref{figS3}(a) \ \ } & \multicolumn{2}{c}{Fig. \ref{figS3}(b) \ \ } & \multicolumn{2}{c}{Fig. \ref{figS3}(c) \ \ } & \multicolumn{2}{c}{Fig. \ref{figS3}(d) \ \ } \\ 
energy (eV)   &\multicolumn{2}{c}  {0.00} & \multicolumn{2}{c}  { 0.26} &  \multicolumn{2}{c} {0.28}           &  \multicolumn{2}{c} {1.27} \\
\hline
&  Ho & Co&           Ho & Co&                Ho & Co&           Ho & Co    \\
spin mag. mom. ($\mu_{\rm B}$) &  3.00 & 1.90        &  4.64  & 1.61  &   4.01 & 2.17        &  2.84 & 1.50 \\
orb. mag. mom. ($\mu_{\rm B}$)&  1.74 & 0.22        &  1.75 & 0.10      & 1.86 & 0.25      & 3.44 & 0.21 \\ 
\hline
\end{tabular}
\caption{Total energies of adsorption geometries of HoCo dimer at 2MgO(100)/Ag(100), relative to the energy of the  most favorable structure shown in Fig.\ref{figS3}(a), as well as spin and orbital magnetic moments of Ho and Co atoms.}
\label{tabS3}
\end{table*}
\end{document}